%% file: cwl-cacm.tex
\documentclass[sigconf,revew,screen,timestamp,nonacm]{acmart}

\usepackage[colorinlistoftodos,prependcaption,textsize=tiny,disable]{todonotes}

\usepackage{xcolor}
\usepackage[normalem]{ulem}

\newcommand{\addition}[1]{{#1}}
\newcommand{\modification}[1]{{#1}}
\newcommand{\deletion}[1]{{}}

\definecolor{OwnAzure}{HTML}{336699}
\definecolor{OwnCerulean}{HTML}{CAE2FE}
\definecolor{OwnOliveGreen}{HTML}{556B2F}
\definecolor{KamPurple}{HTML}{907C97}
\newcommand{\todorone}[1]{\todo[color=orange!40]{R1.#1}}
\newcommand{\todortwo}[1]{\todo[linecolor=white,color=KamPurple!40]{R2.#1}}
\newcommand{\todorthree}[1]{\todo[color=OwnAzure!40]{R3.#1}}

\newcommand{\contributor}[3]
{\normalsize\href{#1}{#2} \small(#3)\normalsize}

\begin{document}

\title{Methods Included: Standardizing Computational Reuse and Portability
with the Common Workflow Language}

\author{Michael R. Crusoe}
\orcid{0000-0002-2961-9670}
\affiliation{%
  \institution{VU Amsterdam}
  \department{\addition{Department of Computer Science}}
  \streetaddress{De Boelelaan 1111}
  \city{Amsterdam}
  \country{The Netherlands}
  \postcode{1081 HV}
}
\affiliation{%
  \institution{Software Freedom Conservancy}
  \department{Common Workflow Language project}
  \streetaddress{137 MONTAGUE ST STE 380}
  \city{Brooklyn}
  \state{NY}
  \country{USA}
  \postcode{11201-3548}
}
\email{mrc@commonwl.org}

\author{Sanne Abeln}
\orcid{0000-0002-2779-7174}
\affiliation{%
  \institution{VU Amsterdam}
  \department{Department of Computer Science}
  \streetaddress{De Boelelaan 1111}
  \city{Amsterdam}
  \country{The Netherlands}
  \postcode{1081 HV}
}
\email{s.abeln@vu.nl}

\author{Alexandru Iosup}
\orcid{0000-0001-8030-9398}
\affiliation{%
  \institution{VU Amsterdam}
  \department{Department of Computer Science}
  \streetaddress{De Boelelaan 1111}
  \city{Amsterdam}
  \country{The Netherlands}
  \postcode{1081 HV}
}
\email{a.iosup@vu.nl}

\author{Peter Amstutz}
\orcid{0000-0003-3566-7705}
\affiliation{%
  \institution{Curii \modification{Corporation}}
  \streetaddress{212 Elm St. 3rd Floor}
  \city{Sommerville}
  \state{MA}
  \country{USA}
  \postcode{02144-2959}
}
\email{peter.amstutz@curii.com}

\author{John Chilton}
\orcid{0000-0002-6794-0756}
\affiliation{%
  \institution{Pennsylvania State University}
  \department{Department of Biochemistry and Molecular Biology}
  \city{State College}
  \state{PA}
  \country{USA}
  \postcode{16801}
}
\affiliation{%
  \institution{Galaxy Project}
  \country{Multiple Countries}
  }
\email{jmchilton@gmail.com}

\author{\modification{Neboj\v{s}a Tijani\'{c}}}
\orcid{0000-0001-8316-4067}
\affiliation{%
  \institution{\modification{Seven Bridges}}
  \streetaddress{Schrafft’s City Center, 529 Main St, Suite 6610}
  \city{Charlestown}
  \state{MA}
  \country{USA}
  \postcode{02129}
}
\email{boysha@gmail.com}

\author{\modification{Herv\'{e} M\'{e}nager}}
\orcid{0000-0002-7552-1009}
\affiliation{%
  \institution{Institut Pasteur}
  \department{\modification{Hub de Bioinformatique et Biostatistique – D\'{e}partement Biologie Computationnelle}}
  \streetaddress{25-28 Rue du Dr Roux}
  \city{Paris}
  \country{France}
  \postcode{75015}
}
\email{herve.menager@pasteur.fr}

\author{Stian Soiland-Reyes}
\orcid{0000-0001-9842-9718}
\affiliation{%
  \institution{The University of Manchester}
  \department{Department of Computer Science}
  \city{Manchester}
  \country{UK}
}
\affiliation{%
  \institution{Informatics Institute, University of Amsterdam}
  \city{Amsterdam}
  \country{Netherlands}
}
\email{soiland-reyes@manchester.ac.uk}

\author{\addition{Bogdan Gavrilovic}}
\orcid{0000-0003-1550-1716}
\affiliation{%
  \institution{Seven Bridges}
  \streetaddress{Schrafft’s City Center, 529 Main St, Suite 6610}
  \city{Charlestown}
  \state{MA}
  \country{USA}
  \postcode{02129}
}
\email{bogdan.gavrilovic@sbgenomics.com}

\author{Carole Goble}
\orcid{0000-0003-1219-2137}
\affiliation{%
  \institution{The University of Manchester}
  \department{Department of Computer Science}
  \city{Manchester}
  \country{UK}
}
\email{carole.goble@manchester.ac.uk}

\author{The CWL Community}
\affiliation{%
  \institution{Software Freedom Conservancy}
  \department{Common Workflow Language project}
  \streetaddress{137 MONTAGUE ST STE 380}
  \city{Brooklyn}
  \state{NY}
  \country{USA}
  \postcode{11201-3548}
}
\email{common-workflow-language@googlegroups.com}

\renewcommand{\shortauthors}{Crusoe et al.}

\input{cwl-abstract}


\maketitle

\section{Introduction} \label{sec:intro}\label{sec:introduction}
\textit{Computational Workflows} are widely used in data analysis, enabling innovation and decision-making for the modern society. But their growing popularity is also a cause for concern: unless we standardize computational reuse and portability, the use of workflows may end up hampering collaboration. How can we enjoy the common benefits of computational workflows and also eliminate such risks?

\todorone{1}\addition{To answer this general question, we advocate in this work for workflow thinking as a shared way of reasoning across all domains and practitioners, introduce \textbf{Common Workflow Language} (CWL) as a pragmatic set of standards for describing and sharing computational workflows, and discuss the principles around which these standards have become central to a diverse community of users across multiple fields of science and engineering.} \todortwo{13}\addition{This article focuses on an overview of the CWL standards and the CWL project and is complemented by the technical detail available in the CWL standards themselves\footnote{https://w3id.org/cwl/v1.2/}.}

\textit{Workflow thinking} \todortwo{4}\addition{is a form of ``conceptualizing processes as recipes and protocols, structured as [work- or] dataflow graphs with computational steps, and subsequently developing tools and approaches for formalizing, analyzing and communicating these process descriptions''}~\cite{gryk_workflows_2017}. \modification{It introduces an abstraction, the workflow, which} helps decouple expertise in a specific domain, for example of \addition{specific} science or engineering \addition{fields}, from expertise in computing. Derived from workflow thinking, a \textit{computational workflow} describes a process for computing where different parts of the process (the tasks) are inter-dependent, e.g., a task can start processing after its predecessors have (partially) completed and where data flows between tasks.

\deletion{In many domains, workflows include diverse analysis components, written in multiple (different) computer languages, by both end-users and third-parties. Such \emph{polylingual} and multi-party workflows are already common or dominant in data-intensive fields like bioinformatics, image analysis, and radio astronomy; we envision they could bring important benefits to many other domains.}

\deletion{To thread data through analysis tools, domain experts such as bioinformaticians use specialized command-line interfaces and other domains use their own customized frameworks. Workflow engines also help with efficient management of the resources used to run scientific workloads.}

\begin{figure*}
  \centering
 \includegraphics[width=\textwidth]{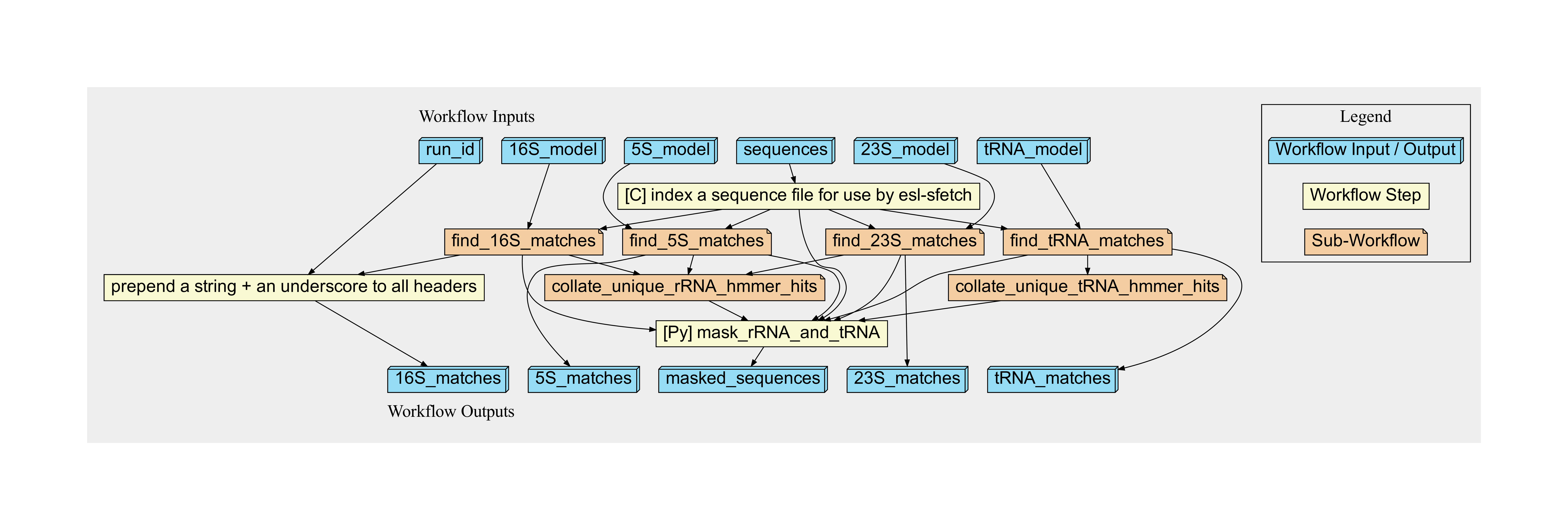}
  \caption{Excerpt from a large microbiome bioinformatics CWL workflow~\cite{mitchell_mgnify_2020}. This part of the workflow \addition{(which is interpretable/executable on its own)} has the aim to match the workflow inputs of genomic sequences to provided sequence-models, which are dispatched to four sub-workflows (e.g., \texttt{find\_16S\_matches}); the sub-workflows not detailed in the figure. The sub-worklow outputs are then collated to identify unique sequence hits, then provided as overall workflow outputs. Arrows define the \modification{connection} between tasks and imply their partial ordering, depicted here as layers of tasks that may execute concurrently.  Workflow steps (e.g., \texttt{mask\_rRNA\_and\_tRNA}) execute command line tools, shown here with indicators for their different programming languages (e.g., \texttt{[Py]} for Python, \texttt{[C]} for the C language). (\modification{Diagram} adapted from \url{https://w3id.org/cwl/view/git/7bb76f33bf40b5cd2604001cac46f967a209c47f/workflows/rna-selector.cwl} \addition{, which was originally retrieved from a corresponding CWL workflow of the EBI Metagenomics project, itself a conversion of the ``rRNASelector''~\cite{lee_rrnaselector_2011} program into a well structured workflow allowing for better parallelization of execution and provenance tracking.})}
  \label{fig:sample_workflow}
  \Description{TODO}
 \end{figure*}

\deletion{The workflow approach helps compose an entire application of these command-line analysis tools: developers build graphical or textual descriptions of how to run these command-line tools, and scientists and engineers connect their inputs and outputs so that the data flows through. An example of a complex workflow problem is metagenomic analysis, for which Figure~\ref{fig:sample_workflow} illustrates a subset (a \textit{sub-workflow}).}

\deletion{In practice, many research and engineering groups use workflows of the kind described in Figure~\ref{fig:sample_workflow}. However, as highlighted in a recently published "Technology Toolbox" article~\cite{perkel_workflow_2019} published in the journal Nature, these groups typically lack the ability to share and collaborate across institutions and infrastructures without costly manual translation of their workflows.}

Currently, many competing systems \todortwo{19}\modification{exist that enable simple workflow execution (\textit{workflow runners}) or offer a comprehensive management of workflows and data (\textit{workflow management systems})}, each with their own syntax or method for describing workflows and infrastructure requirements. This limits computational reuse and portability. In particular, although the data-flows are becoming increasingly more complex, most workflow abstractions do not enable explicit specifications of data-flows, increasing significantly the costs to reuse and port the workflow by third-parties.

We thus identify an important problem for the broad adoption of workflow thinking in practice: although communities \modification{require} polylingual \addition{workflows (workflows that execute tools written in multiple different computer languages)} and multi-party workflows, \textbf{adopting and managing different workflow systems is costly and difficult}. In this work, we propose to tame this complexity through a common abstraction that covers the majority of features used in practice, and is (or can be) implemented in many workflow systems.

In the computational workflow depicted in Figure~\ref{fig:sample_workflow}, practitioners solved the problem by adopting \modification{the CWL standards.} We posit in this work that the \modification{CWL standards} \todortwo{5}\addition{provide the common abstraction that} can help solve the main problems of sharing workflows between institutions and users. \todortwo{6}\addition{CWL achieves this by providing a declarative language that allows expressing computational workflows constructed from diverse software tools, executed each through their command-line interface, with the inputs and outputs of each tool clearly specified and with inputs possibly resulting from the execution of other tools.} We also set out to introduce the CWL standards, with a tri-fold focus: (1)~\modification{the CWL standards} focuses on maintaining a separation of concerns between the description and execution of tools and workflows\addition{, proposing a language that includes only operations commonly used across multiple communities of practice}; (2)~\modification{the CWL standards} support workflow automation, scalability, abstraction, provenance, portability, and reusability; and (3)~the CWL project takes a principled, community-first open-source and open-standard approach which enables this result.

\modification{The CWL standards are} the product of an open and free standards-making community. \modification{While the CWL project} began in the bioinformatics domain\modification{the many \modification{contributors to the CWL project} shaped the standards so that it could be useful anywhere that experiences the problem of ``many tools written in many programming languages by many parties''}. Since the ratification of the first version in 2016, the CWL standards have been used in other fields including hydrology\footnote{\url{https://www.eosc-portal.eu/eoscpilot-science-demonstrator-ewatercycle-switch-fair-data-hydrology}}, radio astronomy\footnote{\url{https://www.eosc-portal.eu/lofar-and-radio-astronomy-community}}, geo-spatial analysis~\cite{simonis_ogc_2020,goncalves_ogc_2020,landry_ogc_2020}, high energy physics~\cite{bell_web-based_2017}, in addition to fast-growing bioinformatics fields like \todortwo{3}\addition{(meta-)}genomics~\cite{mitchell_mgnify_2020} and cancer research~\modification{\cite{lau_cancer_2017}}. \todortwo{2}\todorthree{2}\addition{The CWL standards are featured in US FDA sponsored and adopted IEEE Std 2791\textsuperscript{\texttrademark}-2020 standard~\cite{noauthor_ieee_2020} and the Netherlands' National Plan for Open Science~\cite{van_wezenbeek_nationaal_2017}. A list of free and open-source implementations of the CWL standards are listed in Table \ref{tab:runners}. Additionally there are multiple commercially supported systems that support \modification{the CWL standards} for executing workflows and they are available from vendors such as Curii (Arvados)\footnote{\url{https://www.curii.com/}}, DNAnexus\footnote{\url{https://www.dnanexus.com/}}, IBM (IBM® Spectrum LSF)\footnote{\url{https://github.com/IBMSpectrumComputing/cwlexec}}, Illumina (Illumina Connected Analytics)\footnote{\modification{\url{https://www.illumina.com/products/by-type/informatics-products/connected-analytics.html}}}, and Seven Bridges\footnote{\url{https://www.sevenbridges.com/}}}. The flexibility of \modification{the CWL standards} enabled, for example, rapid collaboration on and prototyping of a COVID-19 public database and analysis resource~\cite{guarracino_covid-19_2020}.

\deletion{CWL could also be useful in computational domains beyond science.} The separation of concerns proposed by \modification{the CWL standards} enable diverse projects, and \modification{can} also benefit engineering and large industrial projects. Likewise, users of \modification{Docker (or other software container technologies)} that distribute analysis tools can use \modification{just the CWL Command Line Tool standard} for providing a structured workflow-independent description of how to run their \modification{tool(s) in the container, what data is required to be provided to the container, and what results to expect and where to find them in the container}.

\textbf{Key Insights [in CACM box]}

Toward computational reuse and portability of polylingual, multi-party workflows, \modification{the CWL project} makes the following contributions:

\begin{enumerate}
\item
  {CWL is a set of standards for describing and sharing computational workflows.}
\item
  {\modification{The CWL standards are} used daily in many science and engineering domains, including by multi-stakeholder teams.}
\item
  {\modification{The CWL standards use} a \textit{declarative syntax}, facilitating polylingual workflow tasks. By being explicit about the \textit{runtime environment} and any use of \textit{software containers}, \modification{the CWL standards} enable \textit{portability} and \textit{reuse}. (See Section~\ref{sec:features}.)}
\item
  {The CWL standards provide a \textit{separation of concerns} between workflow authors and workflow platforms. (More in Section~\ref{sec:open:ecosystem}.)}
\item
  {The CWL standards support critical workflow concepts like automation, scalability, abstraction, provenance, portability, and reusability. (Details in Section~\ref{sec:why}).}
\item
  {\modification{The CWL standards are} developed around core principles of community and shared decision-making, re-use, and zero cost for participants. (Section~\ref{sec:open} details the open standards.)}
\item
  {\modification{The CWL standards are} provided as freely available open standards, supported by a diverse community in collaboration with industry, and is a Free/Open Source Software ecosystem~(see Sidebar~B, Section~\ref{sec:sidebar:b}).}
\end{enumerate}

\section{Background on Workflows and Standards for Workflows}\label{sec:bg}\label{sec:why}

Workflows, and standards-based descriptions thereof, hold the potential to solve key problems in many domains of science and engineering. This section explains why.

\subsection{Why Workflows?}\label{sec:bg:workflows}
\addition{In many domains, workflows include diverse analysis components, written in multiple (different) computer languages, by both end-users and third-parties. Such \emph{polylingual} and multi-party workflows are already common or dominant in data-intensive fields like bioinformatics, image analysis, and radio astronomy; we envision they could bring important benefits to many other domains.

To thread data through analysis tools, domain experts such as bioinformaticians use specialized command-line interfaces~\cite{seemann_ten_2013,georgeson_bionitio_2019} and other domains use their own customized frameworks~\cite{babuji_parsl_2019,berthold_knime_2009}. Workflow engines also help with efficient management of the resources used to run scientific workloads~\cite{deelman_pegasus_2015,couvares_workflow_2007}.

The workflow approach helps compose an entire application of these command-line analysis tools: developers build graphical or textual descriptions of how to run these command-line tools, and scientists and engineers connect their inputs and outputs so that the data flows through. An example of a complex workflow problem is metagenomic analysis, for which Figure~\ref{fig:sample_workflow} illustrates a subset (a \textit{sub-workflow}).

In practice, many research and engineering groups use workflows of the kind described in Figure~\ref{fig:sample_workflow}. However, as highlighted in a recently published ``Technology Toolbox'' article~\cite{perkel_workflow_2019} published in the journal Nature, these groups typically lack the ability to share and collaborate across institutions and infrastructures without costly manual translation of their workflows.}

Using workflow techniques, especially with digital analysis processes, has become quite popular and does not look to be slowing down: one \todortwo{19}{\modification{workflow management system}} recently celebrated its 10,000th citation\footnote{\url{https://galaxyproject.org/blog/2020-08-10k-pubs/}}; and over 29\modification{8} computational \addition{data analysis} workflow systems are known\footnote{\url{https://s.apache.org/existing-workflow-systems}}. \deletion{Why are workflows needed?}

\deletion{Workflows address two key problems.} A process, digital or otherwise, may grow to such complexity that the authors and users of that process have difficulties in understanding its structure, scaling the process, managing the running of the process, and keeping track of what happened in previous enactments of the process. Process dependencies may be undocumented, obfuscated, or otherwise effectively invisible; even an extensively documented process may be difficult to understand by outsiders or newcomers if a common framework or vocabulary is lacking. The need to run the process more frequently or with larger inputs is unlikely to be achieved  by the initial entity (i.e., either script or person) running the process. What seemed once a reasonable manual step (\textit{run this command here and then paste the result there; then call this person for permission}) will, under the pressure of porting and reusing, become a bottleneck. Informal logs (if any) will quickly become unsuitable for answering an organization's need to understand \textit{what} happened, \textit{when}, by \textit{whom}, and to \textit{which} data.

\deletion{A second significant problem is that incomplete method-descriptions are common when computational analysis is reported in academic research~\cite{ivie_reproducibility_2018}. Reproduction, re-use, and replication~\cite{feitelson_repeatability_2015} of these digital methods requires a complete description of what computer applications were used, how exactly they were used, and how they were connected to each other. For precision and interoperability, this description should also be in an appropriate standardized machine-readable format.}

Workflow techniques aim to solve these problems by providing the Abstraction, Scaling, Automation, and Provenance (\textit{A.S.A.P.}) features~\cite{cuevas-vicenttin_scientific_2012}.  Workflow constructs enable a clear abstraction about the \textit{components}, the \textit{relationships} between components, and the \textit{inputs} and \textit{outputs} of the components turning them into well-labeled tools with documented expectations. This abstraction enables \textit{scaling} (execution can be parallelized and distributed), \textit{automation} (the abstraction can be used by a workflow engine to track, plan, and manage execution of tasks), and \textit{provenance} tracking (descriptions of tasks, executors, inputs, outputs; with timestamps, identifiers \todortwo{19}\addition{(\textit{unique names})}, and other logs, can be stored in relation to each other to later answer structured queries).

\subsection{Why Workflow Standards?}\label{sec:bg:standard}

Although workflows are very popular, \modification{prior to the CWL standards every workflow system was incompatible with every other}. This means that those users not using the CWL standards are required to express their computational workflows in a different way every time they have to use another workflow system leading to local success, but global \textit{un}portability. \deletion{Could (the CWL) standards provide a better way?}

The success of workflows is now their biggest drawback: users are locked into a particular vendor, project, and often a particular hardware setup. This hampers sharing and re-use. Even non-academics suffer from this situation, as the lack of standards (or the lack of their adoption) hinders effective collaboration on computational methods within and between companies. Likewise, this \textit{unportability} affects public-private partnerships and the potential for technology transfer from public researchers.

\addition{A second significant problem is that incomplete method descriptions are common when computational analysis is reported in academic research~\cite{ivie_reproducibility_2018}. Reproduction, re-use, and replication~\cite{feitelson_repeatability_2015} of these digital methods requires a complete description of what computer applications were used, how exactly they were used, and how they were connected to each other. For precision and interoperability, this description should also be in an appropriate standardized machine-readable format.}

A standard for sharing and reusing workflows can provide a solution to describing portable, re-usable workflows while also being workflow-engine and vendor-neutral.

\addition{Sharing \textit{workflow descriptions based on standards} also addresses the second problem: the availability of the workflow description provides needed information when sharing; and the quality of the description provided by a structured, standards-based approach is much higher than the current approach of \modification{casual, unstructured, and almost always incomplete} descriptions in scientific reports. Moreover, the operational parts of the description can be provided automated by the workflow management system, rather than by domain experts.}

\todortwo{14}\addition{
While (data) standards are commonly adopted and have become expected for funded projects in knowledge representation fields, the same cannot yet be said about workflows and workflow engines yet.}

\subsection{Sidebar A: Monolingual and Polylingual workflow systems} \label{sec:sidebar:a}

Workflows techniques can be implemented in many ways, i.e., with varying degrees of formalism, which tends to correlate with execution flexibility and features. Typically, whereas the most \textit{informal techniques} require that all processing components are written in the same programming language or are at least callable from the same programming language, the \textit{formal workflow techniques} tend to allow components to be developed in multiple programming languages. 

Among the informal techniques, the \textit{do-it-yourself approach} uses from a particular programming language its built-in capabilities. For example, Python provides a \emph{threading} library, and the Java-based Apache Hadoop~\cite{taylor_overview_2010} provides MapReduce capabilities. To gain more flexibility when working with a particular programming language, \textit{general third-party libraries}, such as \emph{ipyparallel}\footnote{\url{https://pypi.org/project/ipyparallel/}}, can enable remote or distributed execution without having to re-write one's code.

A more explicit workflow structure can be achieved by using a \textit{workflow library} focusing on a specific programming language. For example, in Parsl~\cite{babuji_parsl_2019}, the workflow constructs (``this is a unit of processing'', ``here are the dependencies between the units'') are made explicit and added by the developer to a Python script, to upgrade it to a scalable workflow. (\modification{While we list Parsl here} as an example of a \textbf{monolingual} workflow system, \deletion{although} it also contains explicit support for executing external command-line tools.)

Two approaches can accommodate \textbf{polylingual} workflows where the components are written in more than one programming language, or where the components come from third-parties and the user does not want to or cannot modify them: use of per-language \textit{add-in libraries} or the use of the \textit{Portable Operating System Interface command-line interface (POSIX CLI)}~\cite{the_austin_group_posix1-2008_2008}. The use of per-language add-in libraries entails either explicit function calls (e.g., using \emph{ctypes} in Python to call a C library\footnote{\url{https://docs.python.org/3/library/ctypes.html}}) or the addition of annotations to the user's functions, and requires mapping/restricting to a common cross-language data model.

Essentially all programming languages support the creation of \textbf{POSIX CLIs} \addition{are familiar} to many Linux and macOS users; scripts or binaries which can be invoked on the shell with a set of arguments, reading and writing files, and executed in a separate process. Choosing the POSIX command-line interface as the point of coordination means the connection between components is done by an array of string \textit{arguments} representing program options (including paths to data files) along with a string-based \textit{environment variables} (key-value pairs). \addition{Using the command-line as a coordination interface has the advantage of not needing additional implementation in every programming language, but has the disadvantages of process start-up time and a very simple data model.} \deletion{This command-line option has the advantage of not needing per-language implementation at the expense of a very simple data model (and process start-up costs) which in workflows leads to a tendency for larger granularity of the units of work.} (As a \textit{polylingual} workflow standard, CWL uses the POSIX CLI data model.)

\begin{figure*}
  \centering
  \includegraphics[width=\textwidth]{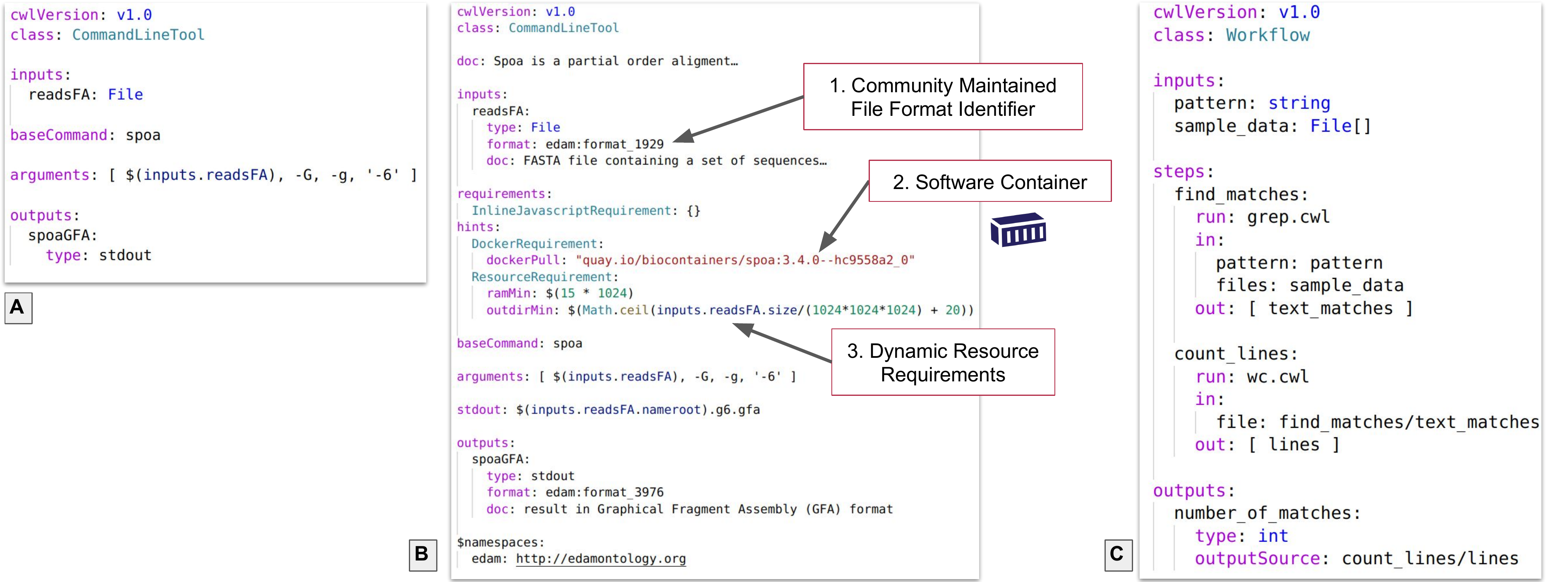}
  \caption{\emph{Example of CWL syntax and progressive enhancement.} (A) and (B) describe the same tool, but (B) is enhanced with additional features:
  human-readable documentation;
  \textit{file format} identifiers \addition{for better}
  validation of workflow connections;
  recommended \textit{software container} image for more reproducible results and easier software installation;
  dynamically specified \textit{resource requirements} to optimize task scheduling and resource usage without manual intervention. The resource requirements are expressed as \textit{hints}. \addition{(C) shows an example of CWL Workflow syntax, where the underlying tool descriptions (``grep.cwl'' and ``wc.cwl'') are in external files for ease of reuse.}}
  \Description{TODO}
  \label{fig:syntax}
 \end{figure*}
 
\section{Features of the Common Workflow Language standards}\label{sec:features}\label{sec:design}

\todorthree{6}\modification{The \textit{Common} Workflow Language standards aim to cover the \textit{common} needs of users and the \textit{common}ly implemented features of workflow runners or platforms. The remainder of this section presents an overview of the CWL features, how they translate to executing workflows in CWL format, and where the CWL standards are not helpful.}

\modification{The CWL standard support} polylingual and multi-party workflows, for which \modification{they enable} computational reuse and portability (see also the CACM Box for main features). To do so, each release of \modification{the CWL standards} has two\todortwo{10}\footnote{The third component, \textit{Schema Salad}, \addition{is only of interest to those who want to parse the syntax of the} schema language that is used to define the syntax of CWL itself.} main components: (1) a standard for describing \textit{command line} tools; and (2) a standard for describing \textit{workflows} that compose such tool descriptions. The goal of the \textbf{CWL Command Line Tool Description Standard}\footnote{\modification{\url{https://w3id.org/cwl/v1.2/CommandLineTool.html}}} is to describe how a particular command line tool works: what are the \textit{inputs} and \textit{parameters} and their types; how to add the correct flags and switches to the \textit{command line} invocation; and where to find the \textit{output files}. 

\modification{The CWL standards} define an \textit{explicit language}, both in syntax, and in its data and execution model. \modification{Its} textual syntax is derived from YAML\footnote{JSON is an acceptable subset of YAML, and common when converting from another format to CWL syntax.}. \modification{This} syntax does not restrict the amount of detail; for example, Figure~\ref{fig:syntax}A depicts a simple example with sparse detail, and Figure~\ref{fig:syntax}B depicts the same example but with the execution augmented with further details. Each \textit{input} to a tool has a name and a type (e.g., File, see label 1 in the figure). 
Authors of tool descriptions are encouraged to include \textit{documentation} and \textit{labels} for all components (i.e., as in Figure~\ref{fig:syntax}B), to enable the automatic generation of helpful visual depictions and even Graphical User Interfaces 
for any given CWL description. \textit{Metadata} about the tool description authors themselves encourages attribution of their efforts. \addition{As shown in Figure~\ref{fig:syntax}B, item 3, these tool descriptions can contain \todortwo{11}\modification{well-defined} \textit{hints} or \addition{mandatory \textit{requirements}} such as which software container to use or how much compute resources are required (memory, number of CPU cores, disk space, and/or the maximum time or deadline to complete the step or entire workflow.)}

\textit{The CWL execution model is explicit}: Each tool's runtime environment is explicit and any \addition{required} elements must be specified by the CWL tool-description author \todortwo{11}\addition{(in contrast to hints, which are optional)}\footnote{\addition{Details on how the CWL Command Line Tool standard specifies that tool executors should setup and control the runtime environment, available  at} \url{https://w3id.org/cwl/v1.2/CommandLineTool.html\#Runtime_environment}\addition{, which also specifies which directories tools are allowed to write to.}}. Each tool invocation uses a separate working directory, populated according to the CWL tool description, e.g., with the input files explicitly specified by the workflow author. \todortwo{17}\addition{Some applications expect particular filenames, directory layouts, and environment variables, and there are additional constructs in the CWL Command Line Tool standard to satisfy their needs.}

\begin{figure*}
  \centering
  \includegraphics[width=0.8\textwidth]{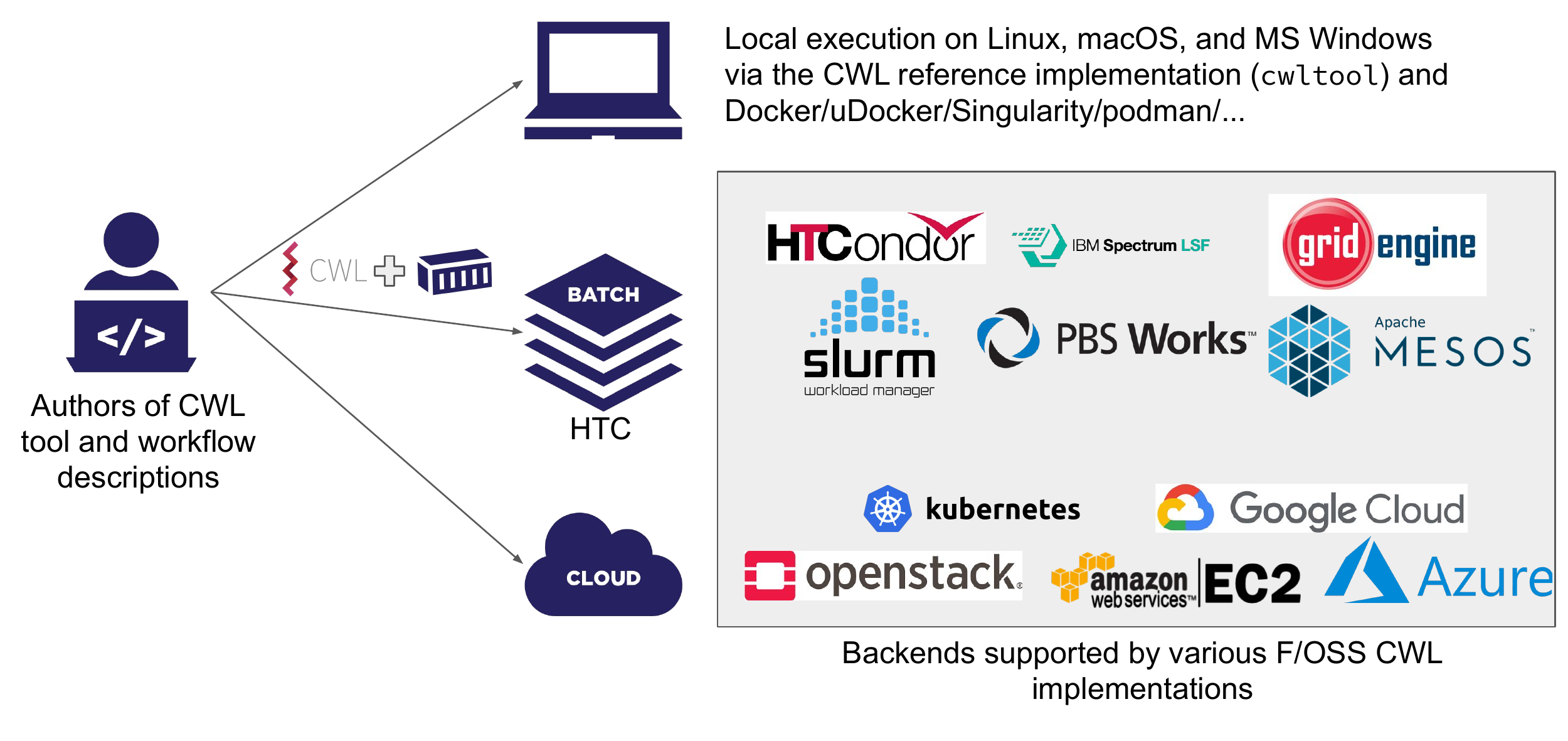}
  \vspace*{-0.5cm}
  \caption{\emph{Example of CWL portability.} The same workflow description runs on the scientist's own laptop or single machine, on any batch production-environment, and on any common public or private cloud. The CWL standards enable execution-portability by being explicit about data locations and execution models.} 
  \label{fig:portability}
  \Description{TODO}
\end{figure*}

\textbf{The explicit runtime model enables portability}, by being explicit about data locations. As Figure~\ref{fig:portability} indicates, this enables execution of CWL workflows on diverse environments as provided by various implementations of the CWL standards: the local environment of the author-scientist (e.g., a single desktop computer, laptop, or workstation), a remote batch production-environment~(e.g., a cluster, an entire datacenter, or even a global multi-datacenter infrastructure), and an on-demand cloud environment.

\modification{The CWL standards explicitly support} the use of \textit{software container} technologies, such as Docker and Singularity, to enable portability of the underlying analysis tools. Figure~\ref{fig:syntax}B, item 2, illustrates the process of pulling a Docker container-image from the \url{Quay.io} registry; then, the workflow engine automates the mounting of files and folders within the container. The container included in the figure has been developed by a trusted author and is commonly used in the bioinformatics field with an expectation its results are reproducible. Indeed, the use of containers can be seen as a confirmation that a tool’s execution is reproducible, when using only its explicitly declared runtime-environment. Similarly, when \todorone{6}\textit{distributed execution} is desired, no changes to the CWL tool-description are needed: because the file or directory inputs are already explicitly defined in the CWL description, the (distributed) \todortwo{19}\modification{workflow runner} can handle (without additional configuration) both job placement and data routing between compute nodes.

\todorone{6}\addition{Via these two features (special handling of data paths; the optional but recommended use of software containers), the CWL standards enables portability (execution “without change”). \todorone{6}Although various factors not controllable by software container technology can affect portability (e.g., variation in the underlying operating system kernel; variation in processor results), in practice the exact same software container and data inputs lead to portability without further adjustment from the user.}

\deletion{The \textit{Common} Workflow Language standards aim to cover the \textit{common} needs of users and the \textit{common}ly implemented features of workflow runners or platforms.} To support features that are \textit{not} in the CWL standards, \modification{the CWL standards define} \textit{extension points} that permit (namespaced) vendor-specific features in explicitly defined ways. If these extensions do not fundamentally change how the tool should operate, then they are \todortwo{11}\modification{added to the \textit{hints}} list and other CWL \addition{compatible} engines can ignore them. However, if the extension is required to properly run the tool being described, e.g., due to the need for some specialized hardware, then the extension is listed under \textit{requirements} and CWL compatible engines can recognize and explicitly declare their inability to execute that CWL description.

The \textbf{CWL Workflow Description Standard}\footnote{\modification{\url{https://w3id.org/cwl/v1.2/Workflow.html}}} builds upon the CWL Command Line Tool Standard: it has the same YAML- or JSON-style syntax, with explicit workflow level inputs, outputs, and documentation~(see Figure~\ref{fig:syntax}\addition{C}). The workflow descriptions consists of a list of \textit{steps}, comprised of CWL CommandLineTools or CWL sub-workflows, each re-exposing their tool's required \textit{inputs}. Inputs \todortwo{7}\addition{for each step} are connected \addition{by referencing the name of either} the common \textit{workflow inputs} or \addition{particular} outputs of other steps. The \textit{workflow outputs} expose selected outputs from workflow steps, making explicit which intermediate step outputs will be returned from the workflow. All connections include identifiers, which CWL document authors are encouraged to name meaningfully, e.g., \texttt{reference\_genome} instead of \texttt{input7}.

CWL workflows form explicit \textit{data flows}, as required for the particular computational analysis. The connectivity between steps defines the partial execution order. Parallel execution of steps is permitted and encouraged whenever multiple steps have all of their inputs satisfied, e.g., in Figure~\ref{fig:sample_workflow}, \texttt{find\_16S\_matches} and \texttt{find\_S5\_matches} are at the same data dependency level and can execute concurrently or sequentially in any order. \todortwo{18}\modification{Additionally, a \textit{scatter} construct allows the repeated execution of a CWL step} \addition{(perhaps overlapping in time, \todorone{6}depending on the resources available) where most of the inputs are the same except for one or more inputs that vary. This is done without requiring the modification of the underlying tool description.} Starting with CWL version 1.2, workflows can also conditionally \textit{skip execution} of a (tool or workflow) step, based upon a specified intermediate input or custom boolean evaluation. Combining these features allows for a flexible \textit{branch} mechanism that allows workflow engines to calculate data dependencies before the workflow starts, and thus retains the predictability of the data flow paradigm.

In contrast to hard-coded approaches that rely on implicit file-paths particular for each workflow, CWL workflows are more \textit{flexible}, \textit{reusable}, and \textit{portable} (which enables scalability). The use in the CWL standards of explicit runtime environments, combined with explicit inputs/outputs to form the data flow, enables step reordering and explicit handling of iterations. The same features enable \textit{scalable} remote execution and, more generally, flexible use of runtime environments. Moreover\addition{,} individual tool definitions from multiple workflows can be reused in any new workflow. 

CWL workflow descriptions are also \textit{future-proof}. Forward compatibility of CWL documents is guaranteed, as each CWL document declares which version of the standards it was written for and minor versions do not alter the required features of the major version. A stand-alone upgrader\footnote{\url{https://pypi.org/project/cwl-upgrader/}} can automatically upgrade CWL documents from one version to the next, and many CWL-aware platforms will internally update user-submitted documents at runtime.

\addition{
\subsection{Execution of workflows in CWL format} \label{sec:execution}
\todorone{1}\todorone{3}CWL is a set of standards, not a particular software product to install, purchase, or rent. The CWL standards need to be implemented to be useful; a list of some implementations of the CWL standards is in Table \ref{tab:runners}. Workflow/tool runners that claim compliance with the CWL standards are allowed significant flexibility in how and where they execute a user's CWL documents as long as they fulfill the requirements written in those documents. For example, they are allowed (and encouraged) to \todorone{6}distribute execution of a workflow across all available computers that can fulfill the resource requirements specified by the user. \todorone{5}Aspects of execution not defined by the CWL standards include (web) APIs for workflow execution and real-time monitoring.

For example details about when a step should be considered ready for execution are available in §4 of CWL Workflow Description standard\footnote{\url{https://w3id.org/cwl/v1.2/Workflow.html\#Workflow}} but once all the inputs are available the exact timing is up to the workflow engine itself. 

\todorone{4}Step execution may result in a temporary or permanent failure, as defined in §4 of CWL Workflow Description standard\footnote{\url{https://w3id.org/cwl/v1.2/Workflow.html\#Workflow_success_and_failure}}. It is up to the workflow engine to control any automatic attempts to recover from failures, e.g., to re-execute a Workflow step. Most workflow engines that implement the CWL standards offer the feature of attempting a number of re-executions, as set by the user, before reporting permanent failure.

\todortwo{4}The CWL community has developed the following optimizations without requiring that users re-write their workflows to benefit:
\begin{enumerate}
\item Automatic streaming of data inputs and outputs instead of waiting for all the data to be downloaded or uploaded (where those data inputs or outputs are marked with ``streamable: true'')
\item Workflow step placement based upon data location~\cite{jiang_tr-19-01_2019}, resource needs, and/or cost of data transfer~\cite{jiang_pivot_2019}
\item \todorone{4}The re-use of the results from previously computed steps, even from a different workflow, as long as the inputs are identical. This can be controlled by the user via the ``WorkReuse'' directive\footnote{\url{https://w3id.org/cwl/v1.2/Workflow.html\#WorkReuse}}.
\end{enumerate}

Real world usage at scale: routinely CWL users and vendors report that they analyze 5000 whole genome sequences in a single workflow execution; one customer of a commercial vendor reported a successful run of a workflow that contained an 8,000-wide step; the entire workflow had 25,000 container executions. By design, the CWL standards do not impose any technical limitations to the size of files processed or to the number of tasks run in parallel. The major scalability bottlenecks are hardware-related --- not having enough machines with enough memory, compute or disk space to process more and more data at a larger scale. As these boundaries move in the future with technological advances, the CWL standards should be able to keep up and not be a cause of limitations.
}

\addition{
\subsection{When is CWL not useful?} \label{sec:limitations}
\todortwo{1}The CWL standards were designed for a particular style of command-line tool based data analysis. Therefore, the following situations are out of scope and not appropriate (or possible) to describe using CWL syntax:

\begin{enumerate}
\item Safe interaction with stateful (web) services
\item Real-time communication between workflow steps
\item Interactions with command line tools beside 1) constructing the command line and making available file inputs (both user provided and synthesized from other inputs just prior to execution) and 2) consuming the output of the tool once its execution is finished, in the form of files created/changed, the POSIX standard output and error streams, and the POSIX exit code of the tool
\item Advanced control-flow techniques beyond conditional steps
\item Runtime workflow graph manipulations: dynamically adding or removing new steps during workflow execution, beyond any predefined conditional step execution tests that are in the original workflow description
\item Workflows that contain cycles: ``repeat this step or sub-workflow a specific number of times'' or ``repeat this step or sub-workflow until a condition is met.''\footnote{Supporting cycles/loops as an optional feature has been suggested for a future version of the CWL standards, but it has yet to be put forth as a formal proposal with a prototype implementation. As a work around, one can launch a CWL workflow from within a workflow system that does support cycles, as documented in the eWaterCycle case study with Cylc~\cite{oliver_workflow_2019}.}
\item Workflows that need particular steps run at or during a specific day/time-frame
\end{enumerate}
}

\section{Open-Source, Open Standards, Open~Community} \label{sec:open}

Given the numerous and diverse set of potential users, implementers, and other stakeholders, we posit that a project like CWL requires \addition{the} combined development of code, standards, and community. Indeed, these requirements were part of the foundational design principles for CWL~(Section~\ref{sec:open:principles}); in the long run, these have fostered free and open source software~(Sidebar~B, in Section~\ref{sec:sidebar:b}), and a vibrant and active ecosystem~(Section~\ref{sec:open:ecosystem}).

\subsection{The CWL Principles} \label{sec:open:principles}

\modification{The CWL project} is based on a set of five principles:

\textbf{Principle 1}: The core of the project is the community of people who care about its goals.

\textbf{Principle 2}: To achieve the best possible results, there should be few, if any, barriers to participation. Specifically, to attract people with diverse experiences and perspectives, there must be no cost to participate.

\textbf{Principle 3}: To enable the \deletion{most good}\addition{best outcomes}, project outputs should be used as people see fit. Thus, the standards themselves must be licensed for reuse, with no acquisition price.

\textbf{Principle 4}: The project must not favor any one company or group over another, but neither should it try to be all things to all people. The community decides.

\textbf{Principle 5}: The concepts and ideas must be tested frequently: tested and functional code is the beginning of evaluating a proposal, not the end.

In time, the CWL project-members learned that this approach is a superset of the OpenStand Principles\footnote{https://open-stand.org/about-us/principles/}, a joint ``Modern Paradigm for Standards'' promoted by the IAB, IEEE, IETF, Internet Society, and W3C. The CWL \addition{project} additions to the OpenStand Principles are: (1) to keep participation free of cost, and (2) the explicit choice of the Apache 2.0 license for all its text, conformance tests, and reference implementations.

\textbf{Necessary and sufficient}: All these principles have proven to be essential for the CWL project. For example, the free cost and open source license~(Principles~2 and~3) has enabled many implementations of the CWL standards, several of which re-use different parts of the reference \todortwo{19}\addition{implementation of the CWL standards (\textit{reference runner})}. Being community-first~(Principle 1) has led to several projects from participants that are outside the CWL standards themselves; the most important contributions have made their way back into the project~(Principle~4).

As part of Principle 5, contributors to the CWL project have developed a suite of conformance tests for each version of the CWL standards. These publicly available tests were critical to \addition{the} CWL \addition{project}'s success: they helped \todortwo{19}\modification{assess} the reference implementation of \modification{the CWL standards themselves}; they provided concrete examples to early adopters; and they enabled the developers and users of production implementations of the CWL standards to confirm their correctness.

\subsection{Sidebar B: The CWL \addition{project} and Free/Open Source Software (F/OSS)}\label{sec:sidebar:b}

\subsubsection{Free and Open Source implementations of \addition{the} CWL \addition{standards}}\footnote{Snapshot of \url{https://www.commonwl.org/\#Implementations}}

By 2021, \modification{the CWL standards have} gained much traction and \modification{are} currently widely supported in practice. In addition to the implementations in Table \ref{tab:runners}, Galaxy~\cite{afgan_galaxy_2018}\footnote{\url{https://github.com/common-workflow-language/galaxy/pull/47}} and Pegasus~\cite{deelman_pegasus_2015}\footnote{\url{https://pegasus.isi.edu/documentation/manpages/pegasus-cwl-converter.html}} have in-development support for \modification{the CWL standards} as well.

Wide adoption benefits from our principles: \modification{The CWL standards} include conformance tests, but the CWL community does not yet test or certify \modification{implementations of the CWL standards}, or specific technology stacks. Instead, \modification{the authors and service provides of workflow runners and workflow management systems} self-certify support for the CWL standards, based on a particular technology configuration they deploy and maintain.

\begin{table}
  \caption{Selected F/OSS workflow runners and platforms that implement the CWL standards.}
  \label{tab:runners}
    \begin{tabular}{ll}
      \toprule
      Implementation & Platform support\\
      \midrule
      \href{https://pypi.org/project/cwltool}{cwltool} & Linux, macOS, Windows (via WSL 2) \\
      & local execution only\\
      \href{https://arvados.org}{Arvados} & in the cloud on AWS, Azure and GCP, \\
      & on premise \& hybrid clusters using Slurm\\
      \href{https://pypi.org/project/toil-cwl-runner}{Toil} & AWS, Azure, GCP, Grid Engine, HTCondor, \\
      & LSF, Mesos, OpenStack, Slurm, PBS/Torque\\
      & also local execution on Linux, macOS, \\
      & MS Windows (via WSL 2)\\
      \href{https://pypi.org/project/cwl-airflow}{CWL-Airflow} & Local execution on Linux, OS X\\
      & or via dedicated Airflow enabled cluster.\\
      \href{https://docs.reana.io/}{REANA} & Kubernetes, \href{https://clouddocs.web.cern.ch/clouddocs/containers/}{CERN OpenStack},\\
      & \href{https://wiki.openstack.org/wiki/Magnum}{OpenStack Magnum}\\
      \bottomrule
\end{tabular}
\end{table}

\subsubsection{F/OSS tools and libraries for working with CWL \addition{format documents}}\footnote{Summarized from \url{https://www.commonwl.org/\#Software_for_working_with_CWL}}\label{sec:sidebar:b:workwith}

CWL plugins for text/code editors exist for Atom, vim, emacs, Visual Studio Code, IntelliJ, gedit, and any text editor that support the ``language server protocol''\todortwo{19} \footnote{\addition{\url{https://microsoft.github.io/language-server-protocol/}}} standard.

There are tools to generate CWL \addition{syntax} from Python \todorthree{7}(via argparse/click \modification{or via functions}), ACD\addition{\footnote{``Ajax Command Definitions'' as produced by the EMBOSS tools: \url{http://emboss.sourceforge.net/developers/acd/}}}, CTD\addition{\footnote{XML-based ``Common Tool Descriptors''~\cite{de_la_garza_desktop_2016} originating in the OpenMS project: \url{https://github.com/WorkflowConversion/CTDSchema}}}, and annotations in IPython Jupyter Notebooks. Libraries to generate and/or read CWL \addition{documents} exist in many languages: Python, Java, R, Go, Scala, \addition{Javascript, Typescript,} and C++.

\subsection{The CWL Ecosystem}\label{sec:open:ecosystem}

Beyond the ratified initial and updated CWL standards released over the last six years, the CWL community has developed many \textit{tools, software libraries, connected specifications}, and has shared CWL descriptions for popular tools. For example, there are software development kits for both Python\footnote{\url{https://pypi.org/project/cwl-utils/}} and Java\footnote{\url{https://github.com/common-workflow-lab/cwljava}} that are generated automatically from the CWL schema; this allows programmers to load, modify, and save CWL documents using an object oriented model that has direct correspondence to the CWL standards themselves. CWL SDKs for other languages are possible by extending the code generation routines\footnote{See the \texttt{*codegen*.py} files in \url{https://pypi.org/project/schema-salad/7.1.20210316164414/}}. (See Sidebar B in Section~\ref{sec:sidebar:b:workwith} for practical details.)

\modification{The CWL standards} support well the acute \textit{need to reuse} (and, correspondingly, \textit{to share}) information on workflow execution, and on authoring and provenance. The CWLProv\footnote{\url{https://w3id.org/cwl/prov/}} prototype was created to show how existing standards~\cite{belhajjame_using_2015,kunze_bagit_2018,missier_w3c_2013} can be combined to represent the provenance of a specific execution of a CWL workflow~\cite{khan_sharing_2019}. Although, to-date, CWLProv has only been implemented in the CWL reference runner, interest is high in additional implementation and further development\deletion{---similarly to what other CWL developments experienced during their first years of existence}.

\deletion{One criterion for evaluating a workflow language is the extent to which it supports a \textit{separation of concerns} between the workflow description authors and the software implementers of the language. This separation of concerns is realized in the CWL standards by using an explicit description approach. This has enabled several optimizations in scheduling (location-based~\cite{jiang_tr-19-01_2019}, cost-based~\cite{jiang_pivot_2019}) and data-organization by researchers and other implementers of the CWL standards.}

\section{Conclusion}\label{sec:conclusion}

The problem of standardizing computational reuse is only increasing in prominence and impact. Addressing this problem, various domains in science, engineering, and commerce have already started to migrate to workflows, but efforts focusing on the portability and even definition of workflows remain scattered. In this work we raise awareness to this problem and propose a community-driven solution.

The Common Workflow Language (CWL) is a family of standards for the description of command line tools and of workflows made from these tools. It includes many features developed in collaboration with the community: support for software containers, resource requirements, workflow-level conditional branching, etc. Built on a foundation of five guiding principles, the CWL project delivers open standards, open-source code, and an open community.

For the past six years, the community around CWL has developed organically\deletion{, with plugins and converters for many systems, along with several production-grade implementations of the standards themselves}. 
\todortwo{1}\todorthree{8}\addition{Organizations looking to write, use, or fund data analysis workflows based upon command-line tools should adopt or even require the CWL standards, because the CWL standards offer a common yet reduced set of capabilities that are both used in practice and implemented in many popular workflow systems. CWL is further valuable because it is supported by a large-scale community, diverse fields have already adopted it, and its adoption is rapidly growing.
Specifically,
\begin{enumerate}
\item By using a reduced set of capabilities, the CWL standards limit the complexity encountered by users when they start to use it, and by operators when they have to implement it. (Feedback from the community indicates these are appreciated.)
\item By using declarative syntax, CWL allows users to specify workflows even if they do not know exactly where the workflows would (later) run. 
\item The CWL project is governed in the public interest and produces freely available open standards. The CWL project itself is not a specific workflow management system, workflow runner, or vendor. This allows potential users, operators, and vendors, to avoid lock-in and be more flexible in the future.
\item By offering standards, the CWL project distinguishes itself especially for the complex interactions that appear in scientific and engineering collaborations. These interactions include defining workflows from many different tools (or steps), sharing workflows, long-term archiving, fulfilling requirements of regulators (e.g., US FDA), making workflow executions auditable and reproducible. (This is particularly useful in cooperative environments, where groups that compete with each other need to collaborate, or in scientific papers where the paper results can be reused very efficiently if the analysis is described in a CWL workflow with publicly available software containers for all steps.)
\item The CWL standards are already implemented, adopted, and used; with many production-grade implementations available as open source and with zero-cost. Thus, the different communities of users of the CWL standards already offer numerous workflow and tool descriptions. (This is akin to how the Python ecosystem of shared libraries, code, and recipes is already helpful.) 
\end{enumerate}
}

To conclude: this is a call for others to embrace workflow thinking and join the CWL community \addition{in creating and sharing portable and complete workflow descriptions.} \todorthree{5}\todorthree{8}\addition{With the CWL standards, the methods are included and ready to (re)use}!

\begin{acks}
The CWL project is immensely grateful to the following self-identified CWL Community members and their contributions to the project:
\contributor{https://orcid.org/0000-0002-2703-8936}{Miguel d'Arcangues Boland}{Software, Bug Reports, Maintenance},
\contributor{}{Alain Domissy}{Conceptualization, Answering Questions, Tools},
\contributor{}{Andrey Kislyuk}{Software, Bug Reports},
\contributor{https://orcid.org/0000-0001-7811-8613}{Brandi Davis-Dusenbery}{Conceptualization, Funding acquisition, Investigation, Project Administration, Resources, Supervision, Business
Development, Event Organizing, Talks},
\contributor{https://orcid.org/0000-0001-9795-7981}{Niels Drost}{Funding Acquisition, Blogposts, Event Organizing, Tutorials, Talks},
\contributor{https://orcid.org/0000-0001-8626-2148}{Robert Finn}{Data acquisition, Funding acquisition, Investigation, Resources, Supervision},
\contributor{https://orcid.org/0000-0001-9292-1533}{Michael Franklin}{Software, Bug Reports, Documentation, Event Organizing, Maintenance, Tools, Answering Questions, Talks},
\contributor{https://orcid.org/0000-0003-1550-1716}{\deletion{Bogdan Gavrilovic}}{\deletion{Conceptualization, Software, Validation, Bug Reports, Blogposts, Maintenance, Tools, Answering Questions, Reviewed Contributions, User Testing}},
\contributor{https://orcid.org/0000-0002-5843-4712}{Manabu Ishii}{Blogposts, Documentation, Examples, Event Organizing, Maintenance, Tools, Answering Questions, Translation, Tutorials, Talks},
\contributor{https://orcid.org/0000-0003-4115-3313}{Sinisa Ivkovic}{Software, Validation, Bug Reports, Tools},
\contributor{https://orcid.org/0000-0002-3468-0652}{Alexander Kanitz}{Software, Business Development, Tools, Talks},
\contributor{https://orcid.org/0000-0002-5044-4692}{Sehrish Kanwal}{Conceptualization, Formal Analysis, Investigation, Software, Validation, Bug Reports, Blogposts, Content, Event Organizing, Maintenance, Answering Questions, Tools, Tutorials, Talks, User Testing},
\contributor{https://orcid.org/0000-0001-9102-5681}{Andrey Kartashov}{Conceptualization, Software, Validation, Examples, Tools, Answering Questions},
\contributor{https://orcid.org/0000-0002-6337-3037}{Farah Khan}{Conceptualization, Formal Analysis, Funding Acquisition, Software},
\contributor{https://orcid.org/0000-0002-6486-3898}{Michael Kotliar}{Software, Validation, Bug Reports, Blogposts, Examples, Maintenance, Answering Questions, Reviewed Contributions, Tools, Talks, User Testing},
\contributor{https://orcid.org/0000-0003-1112-2284}{Folker Meyer}{Tools},
\contributor{https://orcid.org/0000-0002-6388-7353}{Rupert Nash}{Software, Bug Reports, Talks, Videos},
\contributor{https://orcid.org/0000-0003-3705-948X}{Maya Nedeljkovich}{Software, Validation, Visualization, Writing -\/- review \& editing, Bug Reports, Tools, Talks},
\contributor{https://orcid.org/0000-0003-3777-5945}{Tazro Ohta}{Formal Analysis, Funding Acquisition, Resources, Validation, Bug Reports, Blogposts, Content, Documentation, Examples, Event Organizing, Answering Questions, Tools, Translation, Tutorials, Talks, User Testing},
\contributor{https://orcid.org/0000-0002-8021-9162}{Pjotr Prins}{Blogposts, Packaging, Bug Reports},
\contributor{https://orcid.org/0000-0001-9279-9910}{Manvendra Singh}{Software, Blogposts, Packaging, Tools, Reviewed Contributions},
\contributor{https://orcid.org/0000-0002-0959-4429}{Andrey Tovchigrechko}{Conceptualization, Software, Bug Reports},
\contributor{https://orcid.org/0000-0003-3156-2105}{Alan Williams}{Investigation},
\contributor{https://orcid.org/0000-0002-6130-1021}{Denis Yuen}{Software, Bug Reports, Documentation, Tools},
\contributor{https://orcid.org/0000-0002-0415-9655}{Alexander (Sasha) Wait Zaranek}{Conceptualization, Funding Acquisition},
\contributor{https://orcid.org/0000-0003-4716-9121}{Sarah Wait Zaranek}{Conceptualization, Funding Acquisition, Project Administration, Resources, Software, Accessibility, Bug Reports, Business Development, Content, Examples, Event Organizing, Answering Questions, Tutorials, Talks}.

The contributions to the CWL project by the authors of this paper are:
\contributor{https://orcid.org/0000-0002-2961-9670}{Michael R. Crusoe}{Conceptualization, Funding Acquisition, Investigation, Project Administration, Resources, Software, Supervision, Validation, Writing --- original draft, Bug Reports, Business Development, Content, Documentation, Examples, Event Organizing, Maintenance, Packaging, Answering Questions, Reviewing Contributions, Tutorials, Talks},
\contributor{https://orcid.org/0000-0002-2779-7174}{Sanne Abeln}{Writing --- original draft, and review \& editing, Conceptualization, Supervision, Funding acquisition},
\contributor{https://orcid.org/0000-0001-8030-9398}{Alexandru Iosup}{Writing --- original draft, and review, editing, Conceptualization, Supervision, Funding acquisition},
\contributor{https://orcid.org/0000-0003-3566-7705}{Peter Amstutz}{Conceptualization, Formal Analysis, Funding Acquisition, Methodology, Project Administration, Resources, Software, Supervision, Validation, Visualization, Bug Reports, Business Development, Content, Documentation, Examples, Event Organizing, Maintenance, Packaging, Answering Questions, Reviewed Contributions, Security, Tools, Tutorials, Talks},
\contributor{https://orcid.org/0000-0001-8316-4067}{Nebojša Tijanić}{Conceptualization, Software},
\contributor{https://orcid.org/0000-0002-7552-1009}{Hervé Ménager}{Conceptualization, Funding Acquisition, Investigation, Methodology, Project Administration, Resources, Software, Supervision, Bug Reports, Business Development, Examples, Event Organizing, Maintenance, Tools, Tutorials, Talks},
\contributor{https://orcid.org/0000-0001-9842-9718}{Stian Soiland-Reyes}{Conceptualization, Funding Acquisition, Investigation, Project Administration, Resources, Software, Supervision, Validation, User Testing, Writing – review and editing, Bug Reports, Blogposts, Business Development, Content, Documentation, Examples, Event Organizing, Packaging, Tools, Answering Questions, Reviewed Contributions, Security, Tutorials, Talks, Videos},
\contributor{https://orcid.org/0000-0003-1550-1716}{\addition{Bogdan Gavrilovic}}{\addition{Conceptualization, Software, Validation, Bug Reports, Blogposts, Maintenance, Tools, Answering Questions, Reviewed Contributions, User Testing}},
\contributor{https://orcid.org/0000-0003-1219-2137}{Carole A. Goble}{Conceptualization, Funding Acquisition, Resources, Supervision, Audio, Business Development, Content, Examples, Event Organizing, Tools, Talks, Videos}.

Funding acknowledgements: \grantsponsor{EU}{European Commission}{https://ec.europa.eu/programmes/horizon2020/} grants  
BioExcel-2 (SSR)  \small\grantnum[https://cordis.europa.eu/project/id/823830]{EU}{H2020-INFRAEDI-02-2018 823830}\normalsize,
BioExcel (SSR) \small\grantnum[https://cordis.europa.eu/project/id/675728]{EU}{H2020-EINFRA-2015-1 675728}\normalsize,
EOSC-Life (SSR) \small\grantnum[https://cordis.europa.eu/project/id/824087]{EU}{H2020-INFRAEOSC-2018-2 824087}\normalsize,
EOSCPilot (MRC) \small\grantnum[https://cordis.europa.eu/project/id/739563]{EU}{H2020-INFRADEV-2016-2 739563}\normalsize,
IBISBA 1.0 (SSR) \small\grantnum[https://cordis.europa.eu/project/id/730976]{EU}{H2020-INFRAIA-2017-1-two-stage 730976}\normalsize,
ELIXIR-EXCELERATE (SSR, HM) \small\grantnum[https://cordis.europa.eu/project/id/676559]{EU}{H2020-INFRADEV-1-2015-1 676559}\normalsize,
\addition{ASTERICS (MRC) \small\grantnum[https://cordis.europa.eu/project/id/653477]{EU}{INFRADEV-4-2014-2015 653477}\normalsize. \grantsponsor{ELIXIR}{ELIXIR}, the research infrastructure for life-science data, Interoperability Platform Implementation Study (MRC). \small\grantnum[https://elixir-europe.org/about-us/commissioned-services/cwl-2018]{ELIXIR}{2018-CWL}\normalsize.}  Various universities have also co-sponsored this project; we thank Vrije Universiteit of Amsterdam, the Netherlands, where the first three authors have their primary affiliation.
\end{acks}

\bibliographystyle{ACM-Reference-Format}
\bibliography{references}


\end{document}

%% file: cwl-abstract.tex
\begin{abstract}

Computational Workflows are widely used in data analysis, enabling innovation and decision-making for the modern society. In many domains the analysis components are numerous and written in multiple different computer languages by third parties. These polylingual workflows are common in many industries and dominant in fields such as bioinformatics, image analysis, and radio astronomy.
However, in practice many competing workflow systems exist, severely limiting portability of such workflows, thereby hindering the transfer of such workflows between different systems, between different projects and different settings, leading to vendor lock-ins and limiting their generic re-usability.   
Here we present the Common Workflow Language (CWL) project which produces free and open standards for describing command-line tool based workflows. The CWL standards provide a common but reduced set of abstractions that are both used in practice and implemented in many popular workflow systems. The CWL language is declarative, which allows expressing computational workflows constructed from diverse software tools, executed each through their command-line interface. Being explicit about the runtime environment and any use of software containers enables portability and reuse. The CWL project is not specific to a particular analysis domain, it is community-driven, and it produces consensus-built standards.  
Workflows written according to the CWL standards are a reusable description of that analysis that are runnable on a diverse set of computing environments. These descriptions contain enough information for advanced optimization without additional input from workflow authors.
The CWL standards support polylingual workflows, enabling portability and reuse of such workflows, easing for example scholarly publication, fulfilling regulatory requirements, collaboration in/between academic research and industry, while reducing implementation costs. CWL has been taken up by a wide variety of domains, and industries and support has been implemented in many major workflow systems.


\end{abstract}

%% file: cwl-cacm.bbl

\begin{thebibliography}{32}


\ifx \showCODEN    \undefined \def \showCODEN     #1{\unskip}     \fi
\ifx \showDOI      \undefined \def \showDOI       #1{#1}\fi
\ifx \showISBNx    \undefined \def \showISBNx     #1{\unskip}     \fi
\ifx \showISBNxiii \undefined \def \showISBNxiii  #1{\unskip}     \fi
\ifx \showISSN     \undefined \def \showISSN      #1{\unskip}     \fi
\ifx \showLCCN     \undefined \def \showLCCN      #1{\unskip}     \fi
\ifx \shownote     \undefined \def \shownote      #1{#1}          \fi
\ifx \showarticletitle \undefined \def \showarticletitle #1{#1}   \fi
\ifx \showURL      \undefined \def \showURL       {\relax}        \fi
\providecommand\bibfield[2]{#2}
\providecommand\bibinfo[2]{#2}
\providecommand\natexlab[1]{#1}
\providecommand\showeprint[2][]{arXiv:#2}

\bibitem[\protect\citeauthoryear{??}{noa}{2020}]%
        {noauthor_ieee_2020}
 \bibinfo{year}{2020}\natexlab{}.
\newblock \bibinfo{booktitle}{\emph{{IEEE} {Standard} for {Bioinformatics}
  {Analyses} {Generated} by {High}-{Throughput} {Sequencing} ({HTS}) to
  {Facilitate} {Communication}}}.
\newblock \bibinfo{type}{{T}echnical {R}eport} 2791-2020.
  \bibinfo{pages}{1--16} pages.
\newblock
\urldef\tempurl%
\url{https://doi.org/10.1109/IEEESTD.2020.9094416}
\showURL{%
\tempurl}
\newblock
\shownote{Conference Name: IEEE Std 2791-2020.}


\bibitem[\protect\citeauthoryear{Afgan, Baker, Batut, van den Beek, Bouvier,
  Čech, Chilton, Clements, Coraor, Grüning, Guerler, Hillman-Jackson,
  Hiltemann, Jalili, Rasche, Soranzo, Goecks, Taylor, Nekrutenko, and
  Blankenberg}{Afgan et~al\mbox{.}}{2018}]%
        {afgan_galaxy_2018}
\bibfield{author}{\bibinfo{person}{Enis Afgan}, \bibinfo{person}{Dannon Baker},
  \bibinfo{person}{Bérénice Batut}, \bibinfo{person}{Marius van den Beek},
  \bibinfo{person}{Dave Bouvier}, \bibinfo{person}{Martin Čech},
  \bibinfo{person}{John Chilton}, \bibinfo{person}{Dave Clements},
  \bibinfo{person}{Nate Coraor}, \bibinfo{person}{Björn~A Grüning},
  \bibinfo{person}{Aysam Guerler}, \bibinfo{person}{Jennifer Hillman-Jackson},
  \bibinfo{person}{Saskia Hiltemann}, \bibinfo{person}{Vahid Jalili},
  \bibinfo{person}{Helena Rasche}, \bibinfo{person}{Nicola Soranzo},
  \bibinfo{person}{Jeremy Goecks}, \bibinfo{person}{James Taylor},
  \bibinfo{person}{Anton Nekrutenko}, {and} \bibinfo{person}{Daniel
  Blankenberg}.} \bibinfo{year}{2018}\natexlab{}.
\newblock \showarticletitle{The {Galaxy} platform for accessible, reproducible
  and collaborative biomedical analyses: 2018 update}.
\newblock \bibinfo{journal}{\emph{Nucleic Acids Research}}
  \bibinfo{volume}{46}, \bibinfo{number}{W1} (\bibinfo{date}{July}
  \bibinfo{year}{2018}), \bibinfo{pages}{W537--W544}.
\newblock
\showISSN{0305-1048}
\urldef\tempurl%
\url{https://doi.org/10.1093/nar/gky379}
\showDOI{\tempurl}


\bibitem[\protect\citeauthoryear{Babuji, Woodard, Li, Katz, Clifford, Kumar,
  Lacinski, Chard, Wozniak, Foster, Wilde, and Chard}{Babuji
  et~al\mbox{.}}{2019}]%
        {babuji_parsl_2019}
\bibfield{author}{\bibinfo{person}{Yadu Babuji}, \bibinfo{person}{Anna
  Woodard}, \bibinfo{person}{Zhuozhao Li}, \bibinfo{person}{Daniel~S. Katz},
  \bibinfo{person}{Ben Clifford}, \bibinfo{person}{Rohan Kumar},
  \bibinfo{person}{Lukasz Lacinski}, \bibinfo{person}{Ryan Chard},
  \bibinfo{person}{Justin~M. Wozniak}, \bibinfo{person}{Ian Foster},
  \bibinfo{person}{Michael Wilde}, {and} \bibinfo{person}{Kyle Chard}.}
  \bibinfo{year}{2019}\natexlab{}.
\newblock \showarticletitle{Parsl: {Pervasive} {Parallel} {Programming} in
  {Python}}. In \bibinfo{booktitle}{\emph{Proceedings of the 28th
  {International} {Symposium} on {High}-{Performance} {Parallel} and
  {Distributed} {Computing}}} \emph{(\bibinfo{series}{{HPDC} '19})}.
  \bibinfo{publisher}{Association for Computing Machinery},
  \bibinfo{address}{New York, NY, USA}, \bibinfo{pages}{25--36}.
\newblock
\showISBNx{978-1-4503-6670-0}
\urldef\tempurl%
\url{https://doi.org/10.1145/3307681.3325400}
\showDOI{\tempurl}


\bibitem[\protect\citeauthoryear{Belhajjame, Zhao, Garijo, Gamble, Hettne,
  Palma, Mina, Corcho, Gómez-Pérez, Bechhofer, Klyne, and Goble}{Belhajjame
  et~al\mbox{.}}{2015}]%
        {belhajjame_using_2015}
\bibfield{author}{\bibinfo{person}{Khalid Belhajjame}, \bibinfo{person}{Jun
  Zhao}, \bibinfo{person}{Daniel Garijo}, \bibinfo{person}{Matthew Gamble},
  \bibinfo{person}{Kristina Hettne}, \bibinfo{person}{Raul Palma},
  \bibinfo{person}{Eleni Mina}, \bibinfo{person}{Oscar Corcho},
  \bibinfo{person}{José~Manuel Gómez-Pérez}, \bibinfo{person}{Sean
  Bechhofer}, \bibinfo{person}{Graham Klyne}, {and} \bibinfo{person}{Carole
  Goble}.} \bibinfo{year}{2015}\natexlab{}.
\newblock \showarticletitle{Using a suite of ontologies for preserving
  workflow-centric research objects}.
\newblock \bibinfo{journal}{\emph{Journal of Web Semantics}}
  \bibinfo{volume}{32} (\bibinfo{date}{May} \bibinfo{year}{2015}),
  \bibinfo{pages}{16--42}.
\newblock
\showISSN{1570-8268}
\urldef\tempurl%
\url{https://doi.org/10.1016/j.websem.2015.01.003}
\showDOI{\tempurl}


\bibitem[\protect\citeauthoryear{Bell, Canali, Grancher, Lamanna, McCance,
  Mato~Vila, Piparo, Moscicki, Pace, Brito Da~Rocha, Simko, Smith, and
  Tejedor~Saavedra}{Bell et~al\mbox{.}}{2017}]%
        {bell_web-based_2017}
\bibfield{author}{\bibinfo{person}{Tim Bell}, \bibinfo{person}{Luca Canali},
  \bibinfo{person}{Eric Grancher}, \bibinfo{person}{Massimo Lamanna},
  \bibinfo{person}{Gavin McCance}, \bibinfo{person}{Pere Mato~Vila},
  \bibinfo{person}{Danilo Piparo}, \bibinfo{person}{Jakub Moscicki},
  \bibinfo{person}{Alberto Pace}, \bibinfo{person}{Ricardo Brito Da~Rocha},
  \bibinfo{person}{Tibor Simko}, \bibinfo{person}{Tim Smith}, {and}
  \bibinfo{person}{Enric Tejedor~Saavedra}.} \bibinfo{year}{2017}\natexlab{}.
\newblock \bibinfo{booktitle}{\emph{Web-based {Analysis} {Services} {Report}}}.
\newblock \bibinfo{type}{{T}echnical {R}eport} CERN-IT-Note-2018-004.
  \bibinfo{institution}{CERN}, \bibinfo{address}{Geneva, Switzerland}.
\newblock
\urldef\tempurl%
\url{http://cds.cern.ch/record/2315331/}
\showURL{%
\tempurl}


\bibitem[\protect\citeauthoryear{Berthold, Cebron, Dill, Gabriel, Kötter,
  Meinl, Ohl, Thiel, and Wiswedel}{Berthold et~al\mbox{.}}{2009}]%
        {berthold_knime_2009}
\bibfield{author}{\bibinfo{person}{Michael~R. Berthold},
  \bibinfo{person}{Nicolas Cebron}, \bibinfo{person}{Fabian Dill},
  \bibinfo{person}{Thomas~R. Gabriel}, \bibinfo{person}{Tobias Kötter},
  \bibinfo{person}{Thorsten Meinl}, \bibinfo{person}{Peter Ohl},
  \bibinfo{person}{Kilian Thiel}, {and} \bibinfo{person}{Bernd Wiswedel}.}
  \bibinfo{year}{2009}\natexlab{}.
\newblock \showarticletitle{{KNIME} - the {Konstanz} information miner: version
  2.0 and beyond}.
\newblock \bibinfo{journal}{\emph{ACM SIGKDD Explorations Newsletter}}
  \bibinfo{volume}{11}, \bibinfo{number}{1} (\bibinfo{date}{Nov.}
  \bibinfo{year}{2009}), \bibinfo{pages}{26--31}.
\newblock
\showISSN{1931-0145}
\urldef\tempurl%
\url{https://doi.org/10.1145/1656274.1656280}
\showDOI{\tempurl}


\bibitem[\protect\citeauthoryear{Couvares, Kosar, Roy, Weber, and
  Wenger}{Couvares et~al\mbox{.}}{2007}]%
        {couvares_workflow_2007}
\bibfield{author}{\bibinfo{person}{Peter Couvares}, \bibinfo{person}{Tevfik
  Kosar}, \bibinfo{person}{Alain Roy}, \bibinfo{person}{Jeff Weber}, {and}
  \bibinfo{person}{Kent Wenger}.} \bibinfo{year}{2007}\natexlab{}.
\newblock \showarticletitle{Workflow {Management} in {Condor}}.
\newblock In \bibinfo{booktitle}{\emph{Workflows for e-{Science}: {Scientific}
  {Workflows} for {Grids}}}, \bibfield{editor}{\bibinfo{person}{Ian~J. Taylor},
  \bibinfo{person}{Ewa Deelman}, \bibinfo{person}{Dennis~B. Gannon}, {and}
  \bibinfo{person}{Matthew Shields}} (Eds.). \bibinfo{publisher}{Springer},
  \bibinfo{address}{London}, \bibinfo{pages}{357--375}.
\newblock
\showISBNx{978-1-84628-757-2}
\urldef\tempurl%
\url{https://doi.org/10.1007/978-1-84628-757-2_22}
\showDOI{\tempurl}


\bibitem[\protect\citeauthoryear{Cuevas-Vicenttín, Dey, Köhler, Riddle, and
  Ludäscher}{Cuevas-Vicenttín et~al\mbox{.}}{2012}]%
        {cuevas-vicenttin_scientific_2012}
\bibfield{author}{\bibinfo{person}{Víctor Cuevas-Vicenttín},
  \bibinfo{person}{Saumen Dey}, \bibinfo{person}{Sven Köhler},
  \bibinfo{person}{Sean Riddle}, {and} \bibinfo{person}{Bertram Ludäscher}.}
  \bibinfo{year}{2012}\natexlab{}.
\newblock \showarticletitle{Scientific {Workflows} and {Provenance}:
  {Introduction} and {Research} {Opportunities}}.
\newblock \bibinfo{journal}{\emph{Datenbank-Spektrum}} \bibinfo{volume}{12},
  \bibinfo{number}{3} (\bibinfo{date}{Nov.} \bibinfo{year}{2012}),
  \bibinfo{pages}{193--203}.
\newblock
\showISSN{1610-1995}
\urldef\tempurl%
\url{https://doi.org/10.1007/s13222-012-0100-z}
\showDOI{\tempurl}


\bibitem[\protect\citeauthoryear{de~la Garza, Veit, Szolek, Röttig, Aiche,
  Gesing, Reinert, and Kohlbacher}{de~la Garza et~al\mbox{.}}{2016}]%
        {de_la_garza_desktop_2016}
\bibfield{author}{\bibinfo{person}{Luis de~la Garza}, \bibinfo{person}{Johannes
  Veit}, \bibinfo{person}{Andras Szolek}, \bibinfo{person}{Marc Röttig},
  \bibinfo{person}{Stephan Aiche}, \bibinfo{person}{Sandra Gesing},
  \bibinfo{person}{Knut Reinert}, {and} \bibinfo{person}{Oliver Kohlbacher}.}
  \bibinfo{year}{2016}\natexlab{}.
\newblock \showarticletitle{From the desktop to the grid: scalable
  bioinformatics via workflow conversion}.
\newblock \bibinfo{journal}{\emph{BMC Bioinformatics}} \bibinfo{volume}{17},
  \bibinfo{number}{1} (\bibinfo{date}{March} \bibinfo{year}{2016}),
  \bibinfo{pages}{127}.
\newblock
\showISSN{1471-2105}
\urldef\tempurl%
\url{https://doi.org/10.1186/s12859-016-0978-9}
\showDOI{\tempurl}


\bibitem[\protect\citeauthoryear{Deelman, Vahi, Juve, Rynge, Callaghan,
  Maechling, Mayani, Chen, Ferreira~da Silva, Livny, and Wenger}{Deelman
  et~al\mbox{.}}{2015}]%
        {deelman_pegasus_2015}
\bibfield{author}{\bibinfo{person}{Ewa Deelman}, \bibinfo{person}{Karan Vahi},
  \bibinfo{person}{Gideon Juve}, \bibinfo{person}{Mats Rynge},
  \bibinfo{person}{Scott Callaghan}, \bibinfo{person}{Philip~J. Maechling},
  \bibinfo{person}{Rajiv Mayani}, \bibinfo{person}{Weiwei Chen},
  \bibinfo{person}{Rafael Ferreira~da Silva}, \bibinfo{person}{Miron Livny},
  {and} \bibinfo{person}{Kent Wenger}.} \bibinfo{year}{2015}\natexlab{}.
\newblock \showarticletitle{Pegasus, a workflow management system for science
  automation}.
\newblock \bibinfo{journal}{\emph{Future Generation Computer Systems}}
  \bibinfo{volume}{46} (\bibinfo{date}{May} \bibinfo{year}{2015}),
  \bibinfo{pages}{17--35}.
\newblock
\showISSN{0167-739X}
\urldef\tempurl%
\url{https://doi.org/10.1016/j.future.2014.10.008}
\showDOI{\tempurl}


\bibitem[\protect\citeauthoryear{Feitelson}{Feitelson}{2015}]%
        {feitelson_repeatability_2015}
\bibfield{author}{\bibinfo{person}{Dror~G. Feitelson}.}
  \bibinfo{year}{2015}\natexlab{}.
\newblock \showarticletitle{From {Repeatability} to {Reproducibility} and
  {Corroboration}}.
\newblock \bibinfo{journal}{\emph{ACM SIGOPS Operating Systems Review}}
  \bibinfo{volume}{49}, \bibinfo{number}{1} (\bibinfo{date}{Jan.}
  \bibinfo{year}{2015}), \bibinfo{pages}{3--11}.
\newblock
\showISSN{0163-5980}
\urldef\tempurl%
\url{https://doi.org/10.1145/2723872.2723875}
\showDOI{\tempurl}


\bibitem[\protect\citeauthoryear{Georgeson, Syme, Sloggett, Chung, Dashnow,
  Milton, Lonsdale, Powell, Seemann, and Pope}{Georgeson et~al\mbox{.}}{2019}]%
        {georgeson_bionitio_2019}
\bibfield{author}{\bibinfo{person}{Peter Georgeson}, \bibinfo{person}{Anna
  Syme}, \bibinfo{person}{Clare Sloggett}, \bibinfo{person}{Jessica Chung},
  \bibinfo{person}{Harriet Dashnow}, \bibinfo{person}{Michael Milton},
  \bibinfo{person}{Andrew Lonsdale}, \bibinfo{person}{David Powell},
  \bibinfo{person}{Torsten Seemann}, {and} \bibinfo{person}{Bernard Pope}.}
  \bibinfo{year}{2019}\natexlab{}.
\newblock \showarticletitle{Bionitio: demonstrating and facilitating best
  practices for bioinformatics command-line software}.
\newblock \bibinfo{journal}{\emph{GigaScience}} \bibinfo{volume}{8},
  \bibinfo{number}{giz109} (\bibinfo{date}{Sept.} \bibinfo{year}{2019}).
\newblock
\showISSN{2047-217X}
\urldef\tempurl%
\url{https://doi.org/10.1093/gigascience/giz109}
\showDOI{\tempurl}


\bibitem[\protect\citeauthoryear{Gonçalves}{Gonçalves}{2020}]%
        {goncalves_ogc_2020}
\bibfield{author}{\bibinfo{person}{Pedro Gonçalves}.}
  \bibinfo{year}{2020}\natexlab{}.
\newblock \bibinfo{booktitle}{\emph{{OGC} {Earth} {Observations} {Applications}
  {Pilot}: {Terradue} {Engineering} {Report}}}.
\newblock \bibinfo{type}{{OGC} {Public} {Engineering} {Report}} OGC 20-042.
  \bibinfo{institution}{Open Geospatial Consortium}.
\newblock
\urldef\tempurl%
\url{http://docs.opengeospatial.org/per/20-042.html}
\showURL{%
\tempurl}


\bibitem[\protect\citeauthoryear{Group}{Group}{2008}]%
        {the_austin_group_posix1-2008_2008}
\bibfield{author}{\bibinfo{person}{The~Austin Group}.}
  \bibinfo{year}{2008}\natexlab{}.
\newblock \bibinfo{booktitle}{\emph{{POSIX}.1-2008 ({IEEE} {Std} 1003.1™-2008
  and {The} {Open} {Group} {Technical} {Standard} {Base} {Specifications},
  {Issue} 7)}}.
\newblock \bibinfo{type}{{T}echnical {R}eport}. \bibinfo{institution}{Austin
  Group}.
\newblock
\urldef\tempurl%
\url{https://pubs.opengroup.org/onlinepubs/9699919799.2008edition/}
\showURL{%
\tempurl}


\bibitem[\protect\citeauthoryear{Gryk and Ludäscher}{Gryk and
  Ludäscher}{2017}]%
        {gryk_workflows_2017}
\bibfield{author}{\bibinfo{person}{Michael~R. Gryk} {and}
  \bibinfo{person}{Bertram Ludäscher}.} \bibinfo{year}{2017}\natexlab{}.
\newblock \showarticletitle{Workflows and {Provenance}: {Toward} {Information}
  {Science} {Solutions} for the {Natural} {Sciences}}.
\newblock \bibinfo{journal}{\emph{Library trends}} \bibinfo{volume}{65},
  \bibinfo{number}{4} (\bibinfo{year}{2017}), \bibinfo{pages}{555--562}.
\newblock
\showISSN{0024-2594}
\urldef\tempurl%
\url{https://doi.org/10.1353/lib.2017.0018}
\showDOI{\tempurl}


\bibitem[\protect\citeauthoryear{Guarracino, Amstutz, Liener, Crusoe, Novak,
  Garrison, Ohta, Munyoki, Welter, Wait~Zaranek, Wait~Zaranek, and
  Prins}{Guarracino et~al\mbox{.}}{2020}]%
        {guarracino_covid-19_2020}
\bibfield{author}{\bibinfo{person}{Andrea Guarracino}, \bibinfo{person}{Peter
  Amstutz}, \bibinfo{person}{Thomas Liener}, \bibinfo{person}{Michael Crusoe},
  \bibinfo{person}{Adam Novak}, \bibinfo{person}{Erik Garrison},
  \bibinfo{person}{Tazro Ohta}, \bibinfo{person}{Bonface Munyoki},
  \bibinfo{person}{Danielle Welter}, \bibinfo{person}{Sarah Wait~Zaranek},
  \bibinfo{person}{Alexander~(Sasha) Wait~Zaranek}, {and}
  \bibinfo{person}{Pjotr Prins}.} \bibinfo{year}{2020}\natexlab{}.
\newblock \bibinfo{title}{{COVID}-19 {PubSeq}: {Public} {SARS}-{CoV}-2
  {Sequence} {Resource}}.
\newblock
\newblock
\urldef\tempurl%
\url{https://bcc2020.sched.com/event/coLw/covid-19-pubseq-public-sars-cov-2-sequence-resource}
\showURL{%
\tempurl}


\bibitem[\protect\citeauthoryear{Ivie and Thain}{Ivie and Thain}{2018}]%
        {ivie_reproducibility_2018}
\bibfield{author}{\bibinfo{person}{Peter Ivie} {and} \bibinfo{person}{Douglas
  Thain}.} \bibinfo{year}{2018}\natexlab{}.
\newblock \showarticletitle{Reproducibility in {Scientific} {Computing}}.
\newblock \bibinfo{journal}{\emph{Comput. Surveys}} \bibinfo{volume}{51},
  \bibinfo{number}{3} (\bibinfo{date}{July} \bibinfo{year}{2018}),
  \bibinfo{pages}{63:1--63:36}.
\newblock
\showISSN{0360-0300}
\urldef\tempurl%
\url{https://doi.org/10.1145/3186266}
\showDOI{\tempurl}


\bibitem[\protect\citeauthoryear{Jiang, Castillo, and Ahalt}{Jiang
  et~al\mbox{.}}{2019a}]%
        {jiang_tr-19-01_2019}
\bibfield{author}{\bibinfo{person}{Fan Jiang}, \bibinfo{person}{Claris
  Castillo}, {and} \bibinfo{person}{Stan Ahalt}.}
  \bibinfo{year}{2019}\natexlab{a}.
\newblock \bibinfo{booktitle}{\emph{{TR}-19-01: {A} {Cloud}-{Agnostic}
  {Framework} for {Geo}-{Distributed} {Data}-{Intensive} {Applications}}}.
\newblock \bibinfo{type}{Technical {Report}} TR-19-01.
  \bibinfo{institution}{RENCI, University of North Carolina at Chapel Hill}.
  \bibinfo{pages}{10} pages.
\newblock
\urldef\tempurl%
\url{https://renci.org/technical-reports/tr-19-01/}
\showURL{%
\tempurl}


\bibitem[\protect\citeauthoryear{Jiang, Ferriter, and Castillo}{Jiang
  et~al\mbox{.}}{2019b}]%
        {jiang_pivot_2019}
\bibfield{author}{\bibinfo{person}{Fan Jiang}, \bibinfo{person}{Kyle Ferriter},
  {and} \bibinfo{person}{Claris Castillo}.} \bibinfo{year}{2019}\natexlab{b}.
\newblock \bibinfo{booktitle}{\emph{{PIVOT}: {Cost}-{Aware} {Scheduling} of
  {Data}-{Intensive} {Applications} in a {Cloud}-{Agnostic} {System}}}.
\newblock \bibinfo{type}{Technical {Report}} TR-19-02.
  \bibinfo{institution}{RENCI, University of North Carolina at Chapel Hill}.
  \bibinfo{pages}{8} pages.
\newblock
\urldef\tempurl%
\url{https://renci.org/technical-reports/tr-19-02/}
\showURL{%
\tempurl}


\bibitem[\protect\citeauthoryear{Khan, Soiland-Reyes, Sinnott, Lonie, Goble,
  and Crusoe}{Khan et~al\mbox{.}}{2019}]%
        {khan_sharing_2019}
\bibfield{author}{\bibinfo{person}{Farah~Zaib Khan}, \bibinfo{person}{Stian
  Soiland-Reyes}, \bibinfo{person}{Richard~O. Sinnott}, \bibinfo{person}{Andrew
  Lonie}, \bibinfo{person}{Carole Goble}, {and} \bibinfo{person}{Michael~R.
  Crusoe}.} \bibinfo{year}{2019}\natexlab{}.
\newblock \showarticletitle{Sharing interoperable workflow provenance: {A}
  review of best practices and their practical application in {CWLProv}}.
\newblock \bibinfo{journal}{\emph{GigaScience}} \bibinfo{volume}{8},
  \bibinfo{number}{11} (\bibinfo{date}{Nov.} \bibinfo{year}{2019}).
\newblock
\urldef\tempurl%
\url{https://doi.org/10.1093/gigascience/giz095}
\showDOI{\tempurl}


\bibitem[\protect\citeauthoryear{Kunze, Littman, Madden, Scancella, and
  Adams}{Kunze et~al\mbox{.}}{2018}]%
        {kunze_bagit_2018}
\bibfield{author}{\bibinfo{person}{J. Kunze}, \bibinfo{person}{J. Littman},
  \bibinfo{person}{E. Madden}, \bibinfo{person}{J. Scancella}, {and}
  \bibinfo{person}{C. Adams}.} \bibinfo{year}{2018}\natexlab{}.
\newblock \bibinfo{booktitle}{\emph{The {BagIt} {File} {Packaging} {Format}
  ({V1}.0)}}.
\newblock \bibinfo{type}{{RFC}} 8493. \bibinfo{institution}{RFC Editor}.
\newblock
\urldef\tempurl%
\url{https://www.rfc-editor.org/info/rfc8493}
\showURL{%
\tempurl}
\newblock
\shownote{ISSN: 2070-1721.}


\bibitem[\protect\citeauthoryear{Landry}{Landry}{2020}]%
        {landry_ogc_2020}
\bibfield{author}{\bibinfo{person}{Tom Landry}.}
  \bibinfo{year}{2020}\natexlab{}.
\newblock \bibinfo{booktitle}{\emph{{OGC} {Earth} {Observation} {Applications}
  {Pilot}: {CRIM} {Engineering} {Report}}}.
\newblock \bibinfo{type}{{OGC} {Public} {Engineering} {Report}} OGC 20-045.
  \bibinfo{institution}{Open Geospatial Consortium}.
\newblock
\urldef\tempurl%
\url{http://docs.opengeospatial.org/per/20-045.html}
\showURL{%
\tempurl}


\bibitem[\protect\citeauthoryear{Lau, Lehnert, Sethi, Malhotra, Kaushik, Onder,
  Groves-Kirkby, Mihajlovic, DiGiovanna, Srdic, Bajcic, Radenkovic, Mladenovic,
  Krstanovic, Arsenijevic, Klisic, Mitrovic, Bogicevic, Kural, and
  Davis-Dusenbery}{Lau et~al\mbox{.}}{2017}]%
        {lau_cancer_2017}
\bibfield{author}{\bibinfo{person}{Jessica~W. Lau}, \bibinfo{person}{Erik
  Lehnert}, \bibinfo{person}{Anurag Sethi}, \bibinfo{person}{Raunaq Malhotra},
  \bibinfo{person}{Gaurav Kaushik}, \bibinfo{person}{Zeynep Onder},
  \bibinfo{person}{Nick Groves-Kirkby}, \bibinfo{person}{Aleksandar
  Mihajlovic}, \bibinfo{person}{Jack DiGiovanna}, \bibinfo{person}{Mladen
  Srdic}, \bibinfo{person}{Dragan Bajcic}, \bibinfo{person}{Jelena Radenkovic},
  \bibinfo{person}{Vladimir Mladenovic}, \bibinfo{person}{Damir Krstanovic},
  \bibinfo{person}{Vladan Arsenijevic}, \bibinfo{person}{Djordje Klisic},
  \bibinfo{person}{Milan Mitrovic}, \bibinfo{person}{Igor Bogicevic},
  \bibinfo{person}{Deniz Kural}, {and} \bibinfo{person}{Brandi
  Davis-Dusenbery}.} \bibinfo{year}{2017}\natexlab{}.
\newblock \showarticletitle{The {Cancer} {Genomics} {Cloud}: {Collaborative},
  {Reproducible}, and {Democratized}—{A} {New} {Paradigm} in {Large}-{Scale}
  {Computational} {Research}}.
\newblock \bibinfo{journal}{\emph{Cancer Research}} \bibinfo{volume}{77},
  \bibinfo{number}{21} (\bibinfo{date}{Oct.} \bibinfo{year}{2017}),
  \bibinfo{pages}{e3--e6}.
\newblock
\showISSN{0008-5472}
\urldef\tempurl%
\url{https://doi.org/10.1158/0008-5472.can-17-0387}
\showDOI{\tempurl}


\bibitem[\protect\citeauthoryear{Lee, Yi, and Chun}{Lee et~al\mbox{.}}{2011}]%
        {lee_rrnaselector_2011}
\bibfield{author}{\bibinfo{person}{Jae-Hak Lee}, \bibinfo{person}{Hana Yi},
  {and} \bibinfo{person}{Jongsik Chun}.} \bibinfo{year}{2011}\natexlab{}.
\newblock \showarticletitle{{rRNASelector}: {A} computer program for selecting
  ribosomal {RNA} encoding sequences from metagenomic and metatranscriptomic
  shotgun libraries}.
\newblock \bibinfo{journal}{\emph{The Journal of Microbiology}}
  \bibinfo{volume}{49}, \bibinfo{number}{4} (\bibinfo{date}{Sept.}
  \bibinfo{year}{2011}), \bibinfo{pages}{689}.
\newblock
\showISSN{1976-3794}
\urldef\tempurl%
\url{https://doi.org/10.1007/s12275-011-1213-z}
\showDOI{\tempurl}


\bibitem[\protect\citeauthoryear{Missier, Belhajjame, and Cheney}{Missier
  et~al\mbox{.}}{2013}]%
        {missier_w3c_2013}
\bibfield{author}{\bibinfo{person}{Paolo Missier}, \bibinfo{person}{Khalid
  Belhajjame}, {and} \bibinfo{person}{James Cheney}.}
  \bibinfo{year}{2013}\natexlab{}.
\newblock \showarticletitle{The {W3C} {PROV} family of specifications for
  modelling provenance metadata}. In \bibinfo{booktitle}{\emph{Proceedings of
  the 16th {International} {Conference} on {Extending} {Database}
  {Technology}}} \emph{(\bibinfo{series}{{EDBT} '13})}.
  \bibinfo{publisher}{Association for Computing Machinery},
  \bibinfo{address}{New York, NY, USA}, \bibinfo{pages}{773--776}.
\newblock
\showISBNx{978-1-4503-1597-5}
\urldef\tempurl%
\url{https://doi.org/10.1145/2452376.2452478}
\showDOI{\tempurl}


\bibitem[\protect\citeauthoryear{Mitchell, Almeida, Beracochea, Boland, Burgin,
  Cochrane, Crusoe, Kale, Potter, Richardson, Sakharova, Scheremetjew,
  Korobeynikov, Shlemov, Kunyavskaya, Lapidus, and Finn}{Mitchell
  et~al\mbox{.}}{2020}]%
        {mitchell_mgnify_2020}
\bibfield{author}{\bibinfo{person}{Alex~L Mitchell}, \bibinfo{person}{Alexandre
  Almeida}, \bibinfo{person}{Martin Beracochea}, \bibinfo{person}{Miguel
  Boland}, \bibinfo{person}{Josephine Burgin}, \bibinfo{person}{Guy Cochrane},
  \bibinfo{person}{Michael~R Crusoe}, \bibinfo{person}{Varsha Kale},
  \bibinfo{person}{Simon~C Potter}, \bibinfo{person}{Lorna~J Richardson},
  \bibinfo{person}{Ekaterina Sakharova}, \bibinfo{person}{Maxim Scheremetjew},
  \bibinfo{person}{Anton Korobeynikov}, \bibinfo{person}{Alex Shlemov},
  \bibinfo{person}{Olga Kunyavskaya}, \bibinfo{person}{Alla Lapidus}, {and}
  \bibinfo{person}{Robert~D Finn}.} \bibinfo{year}{2020}\natexlab{}.
\newblock \showarticletitle{{MGnify}: the microbiome analysis resource in
  2020}.
\newblock \bibinfo{journal}{\emph{Nucleic Acids Research}}
  \bibinfo{volume}{48}, \bibinfo{number}{D1} (\bibinfo{date}{Jan.}
  \bibinfo{year}{2020}), \bibinfo{pages}{D570--D578}.
\newblock
\showISSN{0305-1048}
\urldef\tempurl%
\url{https://doi.org/10.1093/nar/gkz1035}
\showDOI{\tempurl}


\bibitem[\protect\citeauthoryear{Oliver, Shin, Matthews, Sanders, Bartholomew,
  Clark, Fitzpatrick, Haren, Hut, and Drost}{Oliver et~al\mbox{.}}{2019}]%
        {oliver_workflow_2019}
\bibfield{author}{\bibinfo{person}{H. Oliver}, \bibinfo{person}{M. Shin},
  \bibinfo{person}{D. Matthews}, \bibinfo{person}{O. Sanders},
  \bibinfo{person}{S. Bartholomew}, \bibinfo{person}{A. Clark},
  \bibinfo{person}{B. Fitzpatrick}, \bibinfo{person}{R.~v Haren},
  \bibinfo{person}{R. Hut}, {and} \bibinfo{person}{N. Drost}.}
  \bibinfo{year}{2019}\natexlab{}.
\newblock \showarticletitle{Workflow {Automation} for {Cycling} {Systems}:
  {The} {Cylc} {Workflow} {Engine}}.
\newblock \bibinfo{journal}{\emph{Computing in Science Engineering}}
  (\bibinfo{year}{2019}), \bibinfo{pages}{1--1}.
\newblock
\showISSN{1521-9615}
\urldef\tempurl%
\url{https://doi.org/10.1109/MCSE.2019.2906593}
\showDOI{\tempurl}
\newblock
\shownote{00000.}


\bibitem[\protect\citeauthoryear{Perkel}{Perkel}{2019}]%
        {perkel_workflow_2019}
\bibfield{author}{\bibinfo{person}{Jeffrey~M. Perkel}.}
  \bibinfo{year}{2019}\natexlab{}.
\newblock \showarticletitle{Workflow systems turn raw data into scientific
  knowledge}.
\newblock \bibinfo{journal}{\emph{Nature}}  \bibinfo{volume}{573}
  (\bibinfo{date}{Sept.} \bibinfo{year}{2019}), \bibinfo{pages}{149--150}.
\newblock
\urldef\tempurl%
\url{https://doi.org/10.1038/d41586-019-02619-z}
\showDOI{\tempurl}


\bibitem[\protect\citeauthoryear{Seemann}{Seemann}{2013}]%
        {seemann_ten_2013}
\bibfield{author}{\bibinfo{person}{Torsten Seemann}.}
  \bibinfo{year}{2013}\natexlab{}.
\newblock \showarticletitle{Ten recommendations for creating usable
  bioinformatics command line software}.
\newblock \bibinfo{journal}{\emph{GigaScience}} \bibinfo{volume}{2},
  \bibinfo{number}{2047-217X-2-15} (\bibinfo{date}{Dec.} \bibinfo{year}{2013}).
\newblock
\showISSN{2047-217X}
\urldef\tempurl%
\url{https://doi.org/10.1186/2047-217X-2-15}
\showDOI{\tempurl}


\bibitem[\protect\citeauthoryear{Simonis}{Simonis}{2020}]%
        {simonis_ogc_2020}
\bibfield{author}{\bibinfo{person}{Ingo Simonis}.}
  \bibinfo{year}{2020}\natexlab{}.
\newblock \bibinfo{booktitle}{\emph{{OGC} {Earth} {Observation} {Applications}
  {Pilot}: {Summary} {Engineering} {Report}}}.
\newblock \bibinfo{type}{{OGC} {Public} {Engineering} {Report}} OGC 20-073.
  \bibinfo{institution}{Open Geospatial Consortium}.
\newblock
\urldef\tempurl%
\url{https://docs.ogc.org/per/20-073.html}
\showURL{%
\tempurl}


\bibitem[\protect\citeauthoryear{Taylor}{Taylor}{2010}]%
        {taylor_overview_2010}
\bibfield{author}{\bibinfo{person}{Ronald~C. Taylor}.}
  \bibinfo{year}{2010}\natexlab{}.
\newblock \showarticletitle{An overview of the {Hadoop}/{MapReduce}/{HBase}
  framework and its current applications in bioinformatics}.
\newblock \bibinfo{journal}{\emph{BMC Bioinformatics}} \bibinfo{volume}{11},
  \bibinfo{number}{12} (\bibinfo{date}{Dec.} \bibinfo{year}{2010}),
  \bibinfo{pages}{S1}.
\newblock
\showISSN{1471-2105}
\urldef\tempurl%
\url{https://doi.org/10.1186/1471-2105-11-S12-S1}
\showDOI{\tempurl}


\bibitem[\protect\citeauthoryear{van Wezenbeek, Touwen, Versteeg, and van
  Wesenbeeck}{van Wezenbeek et~al\mbox{.}}{2017}]%
        {van_wezenbeek_nationaal_2017}
\bibfield{author}{\bibinfo{person}{W.~J. S.~M. van Wezenbeek},
  \bibinfo{person}{H.~J.~J. Touwen}, \bibinfo{person}{A.~M.~C. Versteeg}, {and}
  \bibinfo{person}{A.~J.~M. van Wesenbeeck}.} \bibinfo{year}{2017}\natexlab{}.
\newblock \bibinfo{booktitle}{\emph{Nationaal plan open science}}.
\newblock \bibinfo{type}{{T}echnical {R}eport}.
  \bibinfo{institution}{Ministerie van Onderwijs, Cultuur en Wetenschap}.
\newblock
\urldef\tempurl%
\url{https://doi.org/10.4233/uuid:9e9fa82e-06c1-4d0d-9e20-5620259a6c65}
\showDOI{\tempurl}


\end{thebibliography}
